\documentclass[prd,superscriptaddress,nofootinbib,notitlepage]{revtex4-2}

\usepackage[utf8]{inputenc}
\usepackage{bm}
\usepackage[cmex10]{amsmath}
\usepackage [utf8]{inputenc}
\usepackage[usenames,dvipsnames]{xcolor}
\usepackage[export]{adjustbox}
\usepackage{tikz}
\usepackage{physics}
\usepackage{amssymb,appendix,dsfont}
\usepackage{graphicx}
\usepackage[colorlinks,hyperindex, 
           bookmarks=true,bookmarksopen=true]{hyperref}
          \hypersetup    {
        colorlinks=true,
       linkcolor=black,
        urlcolor=MidnightBlue,
   }
\newcommand{\kielce}{Institute of Physics, Jan Kochanowski University, ul. Uniwersytecka 7, 25-406, Kielce, Poland}

\newcommand{\Baku}{Composite Materials Research Center, Azerbaijan State Economic University (UNEC), H. Aliyev 135, AZ 1063, Baku, Azerbaijan}

\newcommand{\baku}{Center for Theoretical Physics, Khazar University, Mehseti 41 Street, Baku, AZ1096, Azerbaijan}

\newcommand{\frankfurt}{Institute for Theoretical Physics, J. W. Goethe University,\\
Max-von-Laue-Straße. 1, 60438 Frankfurt am Main, Germany}

\begin{document}

\title{
Scalar field with a time-independent classical source, not trivial after all:\\ from vacuum decay to scattering}

\author{Leonardo Tinti\footnote{
 \href{mailto:dr.leonardo.tinti@gmail.com}{dr.leonardo.tinti@gmail.com}} }\affiliation{\kielce}

\author{Arthur Vereijken\footnote{
\href{mailto:arthur.vereijken@gmail.com}{arthur.vereijken@gmail.com}}}
\affiliation{\kielce}

\author{Shahriyar Jafarzade\footnote{
\href{mailto:shahriyar.jzade@gmail.com}{shahriyar.jzade@gmail.com}}}
\affiliation{\kielce}\affiliation{\Baku}\affiliation{\baku}

\author{Francesco Giacosa\footnote{\href{mailto:francesco.giacosa@gmail.com}{francesco.giacosa@gmail.com}}} 
\affiliation{\kielce}\affiliation{\frankfurt}

\begin{abstract}
\noindent
Historically it has been believed that a time-independent classical source has no effect on the scattering of relativistic uncharged field, in contrast with single particle quantum mechanics. In this work we show that the dynamics is not trivial. We solve exactly for the scattering amplitudes and find that a key ingredient is the production of particles from the unstable vacuum, conceptually similar to the Schwinger mechanism. We compute exactly the probabilities for the vacuum to decay in $n$ particles. The time dependence of such probabilities displays interesting properties such as the quantum Zeno effect and in particular has no regime where the exponential decay law is a good approximation. We show that the trivial scattering found in the past is the byproduct of the adiabatic switching of the interaction. In fact, it is not possible to switch off the interaction (adiabatically or otherwise) at distant times and recover the exact results. Finally, this non trivial vacuum behavior is a source of particle production. We argue that such non-perturbative calculations can be phenomenologically relevant for the production processes that are suppressed at the lower orders in perturbation theory, for instance dilaton production in a medium. 
\end{abstract}

\maketitle

\section{Introduction}
\noindent
The quantum field theory of an uncharged free scalar particle coupled to an external classical source has received much attention during the development of quantum field theory. This is very natural given its simplicity, and it has become a textbook example illustrating concepts in quantum field theory \cite{Peskin:1995ev,Coleman:2018mew,Wentzel:1949, Ticciati:1999qp}. The claim is usually that if the classical source is static in time, the theory has no scattering and the dynamics is trivial. We will take a close look at the arguments used for this claim and give reasoning why it can be disputed. Namely, in this work we compute exactly the time evolution of quantum states, and it turns out that in the specific case of a time-independent source the scattering is not trivial, the vacuum decays in a similar fashion than in the Schwinger mechanism. In fact, this vacuum decay is the key element to solve all the dynamics of the system, whatever the initial state. 
\\
Vacuum decay of scalar fields has been studied extensively by using, for example, thin wall approximations \cite{Coleman:1977py,Accetta:1989zn,Samuel:1991mz,Hamazaki:1995dy,Bezuglov:2018qpq,Ivanov:2022osf,Brown:2017cca}. It is also closely related to quantum mechanical tunneling \cite{Ai:2020vhx}, in particular when it involves the decay of a metastable vacuum. A possible experimental test of false vacuum decay is proposed in \cite{Fialko:2014xba,Fialko:2016ggg}. One important application is the study of the electroweak vacuum of the Higgs potential \cite{ISIDORI2001387, Buttazzo:2013uya, Devoto:2022qen, Andreassen:2017rzq, GuadaEscalona:2021ueq, Salvio:2016mvj}. Namely, due to quantum corrections to the Higgs self-interaction, the minimum, which classically is global, has the possibility of becoming only a local minimum. Whether or not the electroweak vacuum is only a local metastable vacuum depends on the parameters, in particular, the Higgs mass and the top quark mass. Current knowledge states that it is indeed metastable \cite{Buttazzo:2013uya, Devoto:2022qen,Andreassen:2017rzq} with a lifetime longer than the age of the universe, but close enough to stability that further studies and precise measurements are necessary. The inclusion of gravity also has interesting effects on vacuum stability and vice versa, see e.g. \cite{Accetta:1989zn,Khvedelidze:2000cp,Heyl:2001xp,Higuchi:2008tn}. Recently it has been shown that also in the non-interacting case non trivial vacua can be necessary to properly renormalize the expectation values of observables in flat~\cite{Akkelin:2018hpu,Rindori:2021quq}, as well as non-flat spacetime and cosmology~\cite{Becattini:2022bia,becattini2024negative}. Such non-trivial vacuum effects have been shown to be a source of particles to be taken into account in heavy-ion collisions~\cite{Akkelin:2018hpu,Akkelin:2020cfs}.
\\
Furthermore, scalar fields are plentiful in meson physics as well as BSM physics, examples of such being the dilaton, closely related to the scalar condensate in QCD \cite{Migdal:1982jp,Salomone:1980sp}, and the possible inflaton in cosmology. Perturbation of the ground state by a temporary classical external field is also a convenient way to approximately model a physical phenomenon such as the quark-gluon plasma and the dilaton production from its decay.
\\
\\
The vacuum decay process for scalar fields is conceptually analogous to the nonperturbative Schwinger mechanism \cite{Schwinger:1951nm,Greiner:1985raq,Gelis:2015kya, Ai:2020vhx}, in which particles are created from the vacuum when there exists a strong static electric field. Past the conceptual similarity, the results and how they are reached are different. In the Schwinger mechanism there is particle-antiparticle pair production, in contrast to our case, in which a single particle (and indeed an arbitrary number of particles) can be emitted. The particle production does not follow an exponential and at infinite times the particle content is stable. In particular, the survival probability is quadratic at short times, thus allowing for 
the quantum Zeno effect if a sequence of measurements is performed. In fact, the quantum Zeno effect is realized when the time evolution of the system can be slowed down or frozen by repeated measurements at short time intervals and occurs when the short decay at short time is slower than exponential, see e.g.\cite{Misra:1976by,Fonda:1978dk,Itano:1990zz,Koshino:2004rw,Giacosa:2011xa}. It is an important effect in cavity QED \cite{Haroche:2014gla,2008PhRvL.101r0402B} and potentially even in large systems when gravity is involved \cite{Nikolic:2013pza}. In the case of the anti Zeno effect \cite{Giacosa:2011xa,PhysRevLett.86.2699,PhysRevLett.87.040402} the decay is accelerated by repeated measurements with longer time intervals. They are both generic results of the time evolution of quantum mechanics, but in most cases - for example the decay of muons -  an exponential decay is an extremely good approximation experimentally de facto indistinguishable from the full decay law. In the vacuum decay studied in this work this is not the case, as will be shown the exponential decay is at no time scale a good approximation for the decay law.
\\
We proceed as follows. In section \ref{Lagrangian and Hamiltonian} we describe the system. In section \ref{Decay probability} we compute exact relations for the scattering amplitudes and the probabilities of the vacuum decaying to $n$ particles, and in section \ref{sect:vacuum decay} we analyse the vacuum decay, looking at the small time and large time behaviour, compute the probability distribution for specific external sources, and find the behaviour for large volume. We compare findings to previous work and discuss phenomenological applications in \ref{comparison to literature}. Finally, we conclude in section \ref{Conclusion}. Some details of the calculations and formal arguments are left in the appendices.

\section{Setup of the model}
\label{Lagrangian and Hamiltonian}
We examine a free scalar field coupled to a static external field. The simplicity of this model allows us to do exact calculations, with no need for an expansion in Wick or Feynman diagrams. Unless explicitly stated otherwise, everything will be done in the Heisenberg picture. The action of the system can be written in the following way

\begin{equation}\label{Lagrangian}
    {\cal S} = \int d^4 x \; {\cal L}= \int d^4 x \left\{ \frac{1}{2} \partial_\mu \hat \phi(x) \partial^\mu \hat \phi(x) - \frac{1}{2}m^2 \hat \phi^2 (x) + g\rho({\bf x})\hat \phi (x)\right\}.
\end{equation}
Here $\rho(\mathbf{x})$ is a time independent, classical source of the field operator $\hat{\phi}$ that carries dimensions Energy$^{3}$. Moreover, we assume the source is Fourier transformable.

The Euler-Lagrange equation of motion for the system reads

\begin{equation}
    \Box \, \hat \phi(x) + m^2 \hat \phi (x) = g \rho(\bf x).
\end{equation}
Solutions to this equation of motion are given by
\begin{equation}\label{exact}
   \hat \phi(x) = \hat \phi_{0}(x) + i\, g \int_{0}^{t} dy_{0}\int d^3 y \int \frac{d^3p}{(2\pi)^3 2 E_{\bf p}} \left[ \vphantom{\frac{}{}} e^{-ip\cdot(x-y)} - e^{ip\cdot(x-y)} \right]\,\rho(\bf y) \text{ ,}
\end{equation}
where $\hat\phi_{0}(x)$ is a solution to the homogeneous equation of motion, i.e. $ \Box \, \hat \phi_{0}(x) + m^2 \hat \phi_0(x) = 0 $. We can see it is a solution by explicitly applying the Klein-Gordon operator:
\begin{align}
(\Box + m^2) \hat \phi(x) &=  0 + i\, g \int_{0}^{t} dy_{0}\int d^3 y \int \frac{d^3p}{(2\pi)^3 2 E_{\bf p}} \left[ \vphantom{\frac{}{}} (-p^2+m^2) e^{-ip\cdot(x-y)} -(-p^2+m^2) e^{ip\cdot(x-y)} \right]\,\rho(\bf y) \nonumber  \\ &\phantom{= 0 \;} + i\, g \int d^3 y \int \frac{d^3p}{(2\pi)^3 2 E_{\bf p}} \left[ \vphantom{\frac{}{}} (-iE_{\mathbf{p}}) e^{i\mathbf{p}\cdot(\mathbf{x}-\mathbf{y})} -(i E_{\mathbf{p}}) e^{-i\mathbf{p}\cdot(\mathbf{x}-\mathbf{y})} \right]\,\rho(\bf y) 
\\
&= g \int \frac{d^3p}{(2\pi)^3 } e^{i\mathbf{p}\cdot\mathbf{x}}\int d^3 y \; \vphantom{\frac{}{}}  e^{-i\mathbf{p}\cdot\mathbf{y}}\,\rho(\mathbf{y}) =  g \rho(\mathbf{x}). \nonumber
\end{align}
In the second passage we use that the integrals are invariant under $\mathbf{p} \to -\mathbf{p}$, and in the final passage we apply the Fourier and the anti-Fourier transform. The homogeneous part $\phi_{0}(x)$ can be decomposed into positive and negative energy plane waves as:
\begin{equation}
   \hat \phi_{0}(x) = \int \frac{d^3p}{(2\pi)^3 \sqrt{2E_{\bf p}}} \left[ \vphantom{\frac{}{}} a_{\bf p}e^{-i p\cdot x} +a_{\bf p}^\dagger e^{i p\cdot x} \right] \text{ ,}
    \label{homogeneous sol}
\end{equation}
where $a_{\mathbf{p}},a^{\dagger}_{\mathbf{p}}$ are the particle creation and annihilation operators.
The $y_{0}$ integral in Eq. (\ref{exact}) vanishes for $t=0$, so the homogeneous solution corresponds to the full solution at time 0.
\\
Evaluating the $y_{0}$ integral in \eqref{exact} and using the homogeneous solution \eqref{homogeneous sol}, the exact form of the field for $t>0$ can be written in the explicit form
\begin{equation}\label{exact_field_0}
\begin{split}
   \hat \phi(x) &= \int \frac{d^3p}{(2\pi)^3 \sqrt{2E_{\bf p}}} \left[ \vphantom{\frac{}{}} \left( a_{\bf p} + g\frac{e^{i E_{\bf p} t}-1}{\sqrt{2 E_{\bf p}^3}} \int d^3 y \; e^{-i{\bf p}\cdot{\bf y}} \rho(\bf y)\right)e^{-i p\cdot x} \right. \\
    & \qquad \left. + \left( a^\dagger_{\bf p} + g\frac{e^{-i E_{\bf p} t}-1}{\sqrt{2 E_{\bf p}^3}} \int d^3 y \; e^{i{\bf p}\cdot{\bf y}} \rho(\bf y)\right)e^{i p\cdot x} \right].
\end{split}
\end{equation}
Note that the time dependence of the in-homogeneous part is proportional to $\sin(E_{\mathbf{p}}t)/E_{\mathbf{p}}$, which is a nascent delta function. For $t \to \infty$ it approaches the Dirac delta function $\delta(E_{\mathbf{p}})$, which might hint towards a trivial infinite time behaviour, because the mass sets the on-shell energy strictly larger than zero. However for any finite $t$ we can calculate this integral explicitly and find that it is not simply a peak around $E_{\mathbf{p}} = 0,$ but rather it oscillates and is nonzero even for the massive case when $E_{\mathbf{p}} > 0$. We must be careful treating this expression and take the infinite time limit at a later point, which we will see will leave us with non-trivial dynamics.
\\
Calling $h$ and $f$ the functions

\begin{equation}\label{g_f}
\begin{split}
    h({\bf p}) &= g\,\frac{-1}{\sqrt{2E_{\bf p}^3}} \int d^3 y \; e^{-i {\bf p}\cdot {\bf y}}\rho({\bf y}),\\
    f(t,{\bf p}) &= g \,\frac{e^{i E_{\bf p} t}-1}{\sqrt{2E_{\bf p}^3}} \int d^3 y \; e^{-i {\bf p}\cdot {\bf y}}\rho({\bf y}),
    \end{split}
\end{equation}
equation~(\ref{exact_field_0}) can be written in the compact way

\begin{equation}\label{exact_field}
   \hat \phi(x) = \int \frac{d^3 p}{(2\pi)^3 \sqrt{2E_{\bf p}}} \left[  \left(  \vphantom{\frac{}{}} a_{\bf p} +f(t,{\bf p}) \right) e^{-ip\cdot x} + \left( \vphantom{\frac{}{}} a^\dagger_{\bf p} + f^*(t,{\bf p}) \right) e^{ip\cdot x} \right].
\end{equation}
The field can be written in a form similar to \eqref{homogeneous sol}, but with time dependent creation and annihilation operators. From \eqref{exact_field} we can immediately see the form of the time dependent ladder operators 
\begin{equation}
a_{\bf p}(t) =a_{\bf p} + f(t,\mathbf{p}) =a_{\mathbf{p}} + (1-e^{i E_{\bf p} t}) h(\mathbf{p}) \text{ .}   
\end{equation}
{To avoid confusion, we note that the operators $a_{\bf p}(t)$, despite the explicit time dependence, do not evolve with the Heisenberg equation: $a_{\bf p}(t)\neq U^\dagger(t,0)a_{\bf p}(0)U(t,0)$. This is necessary to be consistent with the convention used for the free case~\eqref{homogeneous sol}. In the limit $g\to 0$ the $a_{\bf p}(t)$ reduce to the free $a_{\bf p}$, which have non trivial commutation relations with the free Hamiltonian, but they are constant. As it will be shown and used extensively in Section.~\ref{Decay probability}, they in fact evolve with the interaction picture propagator $U_I= U_0^\dagger U$. }

From this relation we have, by construction, that $a_{\mathbf{p}}(0)$ is just $a_{\mathbf{p}}$ which is the initial condition, so we will omit the time argument at $t = 0$. The field, even in an interacting theory, can always be written in terms of time dependent creation and annihilation operators, see appendix \ref{sec:app_ham}. The free field operators are shifted by a function, and so the expectation value of the particle number operator with respect to the particle vacuum $\ket{0}$ is
\begin{equation}
  N_0(t) = \bra{0}\hat N(t)\ket{0} = \int \frac{d^3 p}{(2\pi)^3 }\bra{0} a^\dagger_{\bf p}(t)a_{\bf p}(t) \ket{0} = \int \frac{d^3 p}{(2\pi)^3 } \abs{f(t,\bf p)}^2 =  2\int \frac{d^3p}{(2\pi)^3} \left( \vphantom{\frac{}{}} 1- {\rm cos}(E_{\bf p}t)  \right) \left|  h({\bf p})\right|^2.
    \label{exp value 1}
\end{equation}
At this point, we note that the number operator $\hat N(t)$ does evolve according to the Heisenberg equation, as it is normal for observables, even if the ladder operators do not. The combination $a^\dagger_{\bf p}(t)a_{\bf p}(t) = U_I^\dagger  \, a^\dagger_{\bf p}\,U_IU_I^\dagger \,a_{\bf p} \, U_I= U_I^\dagger \, a^\dagger_{\bf p} a_{\bf p} \, U_I$ also reads $a^\dagger_{\bf p}(t)a_{\bf p}(t)= U^\dagger \, (U_0 \, a^\dagger_{\bf p}\,U_0^\dagger) \, (U_0 \, a_{\bf p} \, U_0^\dagger ) \, U $. The commutation relations between the ladder operators and the free Hamiltonian are the standard free ones. The phases for the free evolution $U_0\,  a_{\bf p} \, U_0^\dagger = e^{iE_{\bf p}t} a_{\bf p}$ and its Hermitian conjugate exactly compensate, since the creation and annihilation operators act on the same ${\bf p}$, hence the same energy.
The normal-ordered Hamiltonian is computed in Appendix~\ref{sec:app_ham} for a general source; if the source is time independent it turns out the Hamiltonian is also time independent:
\begin{equation}\label{hamiltonian_0}
   \hat H = \int \frac{d^3 p}{(2\pi)^3} \, E_{\bf p} \, \left[  \left(a^\dagger_{\bf p} + h^*({\bf p}) \vphantom{\frac{}{}} \right)\left(a_{\bf p} + h({\bf p})\vphantom{\frac{}{}}\right) -|h({\bf p})|^2 \right] \text{ .}
\end{equation}
As long as the integral
\begin{equation}
    \int \frac{d^3 p}{(2\pi)^3} \, E_{\bf p} |h({\bf p})|^2 = \int \frac{d^3 p}{(2\pi)^3 2 E_{\bf p}^2} \, \left|  \int d^3x \; e^{i{\bf p}\cdot {\bf x}} \rho({\bf x}) \right|^2\in (0,\infty)
\end{equation}
converges, the interacting Hamiltonian corresponds to a finite (and positive) off-set plus a ``free-like'' term, having the same spectrum as the free Hamiltonian, but with the creation (and annihilation) operators shifted by a function. The Hamiltonian, unsurprisingly, has the same spectrum as a free Hamiltonian. The final Hamiltonian is, however, not diagonal for the basis that diagonalizes the free one, since the fields themselves are shifted. This is necessary to have any evolution at all of the observables in the Heisenberg picture of quantum mechanics, since the operators change in time, not the states (except for measurements). The creation and annihilation operators present in the Hamiltonian are different from those in the field as they are shifted by a different factor. After all, the fields are time dependent through the creation and annihilation operators $a_{\mathbf{p}}(t)$, while the Hamiltonian is time independent. This is already a first hint that the Hamiltonian does not have a basis of mutual eigenstates with the momentum or the particle number operator. In appendix \ref{sec:app_ham} more details as well as the general form of the Hamiltonian for a time dependent source are given.

\section{Scattering amplitudes}
\label{Decay probability}
\subsection{General expressions}
Given a generic initial state of the system $|\psi_i\rangle$ measured/prepared at time $t_i = 0$, the probability amplitude and probability to measure the state $|\psi_f\rangle$ at a later time $t>0$ are given by the time-evolution operator of the theory $U(t,0)$ by
\begin{equation}
    a_{i\to f}(t)=\langle \psi_f| U(t,0) |\psi_i\rangle \text{ ,}  \phantom{....}      P_{i\to f}(t) = |a_{i\to f}(t)|^2 \text{ .} 
    \label{eq: probability amplitude}
\end{equation}
The operator $U(t,0)$ is also referred to as the `propagator operator' \cite{Peskin:1995ev,Sakurai:2011zz} (but clearly this expression should not be confused with the Feynman propagator or similar quantities). We will make use of the expression $U(t,0)$ as the evolution-time propagator or propagator `tout court' below.

The probability amplitude simplifies significantly when the Hamiltonian is time independent. For a time independent source, the Hamiltonian is also time independent. When the source depends on time then so will the Hamiltonian, and solving \eqref{eq: probability amplitude} becomes a non-trivial task. Before focusing on the case for the static source, we will solve the dynamics for an arbitrary source, and simplify to a static source later on. One way to tackle the problem is to solve exactly for the Dyson series. However, since there are the full solutions of the Euler-Lagrange equation of motion at all time \eqref{exact_field}, it is possible to avoid altogether solving for the operator $U(t,0)$.
It requires a small detour into some unphysical states and some conceptual considerations over the states and how to recognize them in measurements.
As it is explained in Appendix~\ref{sec:app_ham}, any uncharged scalar field can be decomposed in terms of time-dependent ladder operators, and the explicit form is taken from \eqref{exact_field}. This allows us to use all the free-field concepts of normal ordering and build the particles state. In the appendix \ref{sec:app_ham}, it has been used to build a free field $\phi_0$ that coincides with the interacting one at the time $t_0=0$. Since the fields coincide, the particles states built with the free field are the particle states of the interacting scalar field at $t=0$. In the Heisenberg picture the physical states do not evolve in time, therefore the states at time $t_0=0$ are the physical states at all times.
On the other hand, there is nothing special about fixing $t_0=0$. The very same procedure could have been done at a different time, providing an equally valid base.
\\
As will be shown next, it is possible to use this freedom of choice of the base to exactly solve the dynamics of the system, while bypassing the need to solve for the time-evolution propagator $U(t,0)$. The main ingredient is to notice that the time-dependent ladder operator evolve with the interaction propagator. In general for quantum field theories of scalar fields we can write the annihilation operator as (see appendix \ref{sec:app_ham} for details):

\begin{equation}
    a_{\bf p}(t) = \frac{i}{\sqrt{2E_{\bf p}}} \int d^3 x \; e^{i E_{\bf p}t } \, e^{-i{\bf p}\cdot {\bf x}}\left[ \vphantom{\frac{1}{1}} {\hat \Pi}(t,{\bf x}) - i E_{\bf p} \, {\hat \phi}(t,{\bf x}) \right],
\end{equation}
where $\hat \Pi (t,x) = \partial \cal L / \partial \dot{ \phi}$. 
The hermitian conjugate gives the expression for the creation operator.

The interacting (Heisenberg) fields evolve according with the full propagator operator $U(t,0)$ of the theory, therefore

\begin{equation}
    \begin{split}
        a_{\bf p}(t) &= \frac{i}{\sqrt{2E_{\bf p}}} \int d^3 x \; e^{i E_{\bf p}t } \, e^{-i{\bf p}\cdot {\bf x}}\left[ \vphantom{\frac{1}{1}} U^{-1}(t,0)\,{\hat \Pi}(0,{\bf x})\, U(t,0) - i E_{\bf p} \, U^{-1}(t,0)\,{\hat \phi}(0,{\bf x})\, U(t,0) \right] \\
        &=  U^{-1}(t,0)\left\{\frac{i}{\sqrt{2E_{\bf p}}} \int d^3 x \; e^{i E_{\bf p}t } \, e^{-i{\bf p}\cdot {\bf x}}\left[ \vphantom{\frac{1}{1}}{\hat \Pi}(0,{\bf x}) - i E_{\bf p} \,{\hat \phi}(0,{\bf x}) \right]\right\}  U(t,0)\\
        &=  U^{-1}(t,0)\left\{ \vphantom{\frac{1}{2}}e^{i E_{\bf p}t} \; a_{\bf p} \right\}  U(t,0) \text{ .}   
    \end{split}
\end{equation}
On the other hand, the ladder operators have the same commutation relations with $\hat H_0$ as the free field. That is, with the normal ordered Klein-Gordon Hamiltonian, built with the fields at the same time as the ladder operators. Thus the phase $e^{iE_{\bf p}t}$ can be written in terms of the free propagator $U_0(t,0) = {\rm exp}\{-i\hat H_0 t\}$, leading to
\begin{equation}
    \begin{split}
        a_{\bf p}(t) &=  U^{-1}(t,0)\left\{ \vphantom{\frac{1}{2}}e^{i E_{\bf p}t} \; a_{\bf p} \right\}  U(t,0)\\
        &= U^{-1}(t,0)\left\{ \vphantom{\frac{1}{2}}U_0(t,0) \, a_{\bf p}\, U_0^{-1}(t,0) \right\}  U(t,0) = U_I^{-1}(t,0)\,  a_{\bf p}\, U_I(t,0) \text{ ,}
    \end{split}
\end{equation}
where we recognize the propagator in the interaction picture $U_I = U_0^{-1}U$. The very same arguments hold for the creation operator, since the numerical phase in the time has the opposite sign, as it is required for the opposite sign of the commutation relations with $\hat H_0$ compared to the annihilation (or destruction) operator. Using this fact that the creation and annihilation operators evolve with the interaction picture propagator, it is possible to recognize a continuous set of states labeled by $t$:

\begin{equation}\label{t_states}
    \begin{split}
        |0\rangle_t = U_I^{-1}(t, 0)|0\rangle \quad \Rightarrow \quad a_{\bf p}(t) |0\rangle_t &=\left( \vphantom{\frac{}{}}U_I^{-1}(t, 0)\, a_{\bf p} \, U_I(t, 0) \right) U_I^{-1}(t, 0)|0\rangle \\
        &=U_I^{-1}(t, 0)\left( \vphantom{\frac{}{}} a_{\bf p}|0\rangle \right) = 0, \\
        |{\bf p_{1}},\cdots,{\bf p}_n\rangle_t = U_I^{-1}(t, 0)|{\bf p_{1}},\cdots,{\bf p}_n\rangle \quad \Rightarrow & \\
        \quad a^\dagger_{{\bf p}}(t)\, a_{{\bf p}}(t)\, |{\bf p_{1}},\cdots,{\bf p}_n\rangle_t &= U_I^{-1}(t, 0)\left(\vphantom{\frac{}{}} a^\dagger_{{\bf p}}\, a_{{\bf p}}\, |{\bf p_{1}},\cdots,{\bf p}_n\rangle\right).
    \end{split} 
\end{equation}
They are eigenstates of the number operator at time $t$, and the $|\cdots\rangle$ are just the regular Heisenberg states, eigenstates of the number operator at time $t=0$. It is important to note that these states are not physical states\footnote{They are not normalized to one. However they are a basis for the physical states as usual.}, nor are they states in the Interaction representation, since they are linked by $U_I^{-1}$ instead of the regular $U_I$. They are also not energy eigenstates, as there are no simultaneous eigenstates of the Hamiltonian and the number operator (or equivalently of the Hamiltonian and the momentum operator, see appendix \ref{sec:app_ham} for details). We are still working completely in the Heisenberg picture, and these states are just a useful tool for the calculation. By construction we have
\begin{equation}
    {}_t\langle {\bf p}_1,\cdots ,{\bf p}_n|{\bf k}_1,\cdots {\bf k}_n\rangle =\langle {\bf p}_1,\cdots ,{\bf p}_n|\, U_I(t,0) \,|{\bf k}_1,\cdots {\bf k}_n\rangle.
    \label{t state UI}
\end{equation}
A short note on which states to project on is in order: it might be intuitive to think that, since the basis of particle number eigenstates changes over time, one has to sandwich the propagator between the initial time $n$ particle state and an $n$ particle state at final time. The reason why this should not be done is most easily seen in the Schrödinger picture, where the basis of eigenstates of an operator does not change over time, and one would sandwich in between two states of the same basis. Matrix elements are equivalent between pictures, so it is correct to sandwich the operator between two $t=0$ eigenstates\footnote{Or between two eigenstates for some other time $t=t'$ as long as they are taken at the same time, since the basis spans the Hilbert space for all times all the information is still present in these inner products.}. Thus the regular matrix elements of the full propagator read

\begin{equation}\label{scattering_time_dep}
    \begin{split}
        &\langle {\bf p}_1,\cdots ,{\bf p}_n|\, U(t,0) \,|{\bf k}_1,\cdots {\bf k}_l\rangle = \langle {\bf p}_1,\cdots ,{\bf p}_n|\,U_0(t,0) \, U_0^{-1}(t,0) \,  U(t,0) \,|{\bf k}_1,\cdots {\bf k}_l\rangle \\ 
        & \qquad = \langle {\bf p}_1,\cdots ,{\bf p}_n|\,U_0(t,0) \,  U_I(t,0) \,|{\bf k}_1,\cdots {\bf k}_l\rangle = \left[\prod_{j=1}^n e^{ -iE_{{\bf p}_j}t } \right]\; {}_t\langle {\bf p}_1,\cdots ,{\bf p}_n|{\bf k}_1,\cdots {\bf k}_l\rangle.
    \end{split}
\end{equation}
Here we see more precisely the idea of using the time dependent particle number eigenstates; one projects it without the propagator and with some additional phases. It is possible to use this result to solve for the scattering exactly. Namely, from the definition~\eqref{t_states}, we can recognize that the $t$ states are built just like the regular $n$ particle states, but with the $t$ vacuum and time dependent creation operators:
\begin{equation}
    |{\bf p}_1,\cdots,{\bf p}_n\rangle_t = \sqrt{\frac{\prod_{j=1}^n (2 E_{{\bf p}_j})}{n!}} \; a^\dagger_{{\bf p}_1}(t) \cdots a^\dagger_{{\bf p}_n}(t) \, |0\rangle_t \text{ .}
\end{equation}
The explicit form of the $a_{\bf p}(t)$ in terms of the $a_{\bf p}$ is known exactly from Eq. \eqref{exact_field}, consisting in a shift. In particular, the mixed commutators still give a Dirac delta
\begin{equation}
\label{eq: commutation a(t) a(0)}
    a_{\bf p}(t) = a_{\bf p} + f(t,{\bf p})  \qquad \Rightarrow \qquad \left[ a_{\bf p}(t), a^\dagger_{\bf p}\right] = (2\pi)^3 \delta^3({\bf p} - {\bf q}),
\end{equation}
and the vacua are proportional to a coherent state of each other (a quick overview of coherent states is given in appendix \ref{sec:app_coh})
\begin{equation}\label{prop_to_coh}
    \begin{split}
        & a_{\bf p}(t)\, |0\rangle_t = 0 \quad  \Rightarrow \quad  0 = \left( \vphantom{\frac{}{}} a_{\bf p} + f(t, {\bf p}) \right) |0\rangle_t \quad \Rightarrow \quad  a_{\bf p} |0\rangle_t = - f(t, {\bf p})|0\rangle_t, \\
        & a_{\bf p}\, |0\rangle = 0 \quad  \Rightarrow \quad  0 = \left( \vphantom{\frac{}{}} a_{\bf p}(t) - f(t, {\bf p})\right)|0\rangle \quad \Rightarrow \quad  a_{\bf p}(t) |0\rangle=  f(t, {\bf p})|0\rangle.
    \end{split}
\end{equation}
Applying the latter~\eqref{prop_to_coh} to~\eqref{scattering_time_dep} one almost has the propagator matrix element already
\begin{equation}\label{scattering_time_dep_a}
    \begin{split}
        &\langle {\bf p}_1,\cdots ,{\bf p}_n|\, U(t,0) \,|{\bf k}_1,\cdots {\bf k}_l\rangle = \left[\prod_{j=1}^n e^{ -iE_{{\bf p}_j}t } \right]\; {}_t\langle {\bf p}_1,\cdots ,{\bf p}_n|{\bf k}_1,\cdots {\bf k}_l\rangle \\\\
        &= \sqrt{\frac{\prod_{j=1}^n \left( 2E_{{\bf p}_j} \right)}{n!}}\sqrt{\frac{\prod_{s=1}^l \left( 2E_{{\bf k}_s} \right)}{l!}}\left[\prod_{j=1}^n e^{ -iE_{{\bf p}_j}t } \right]\; {}_t\langle 0|   \, a_{{\bf p}_1}(t)\cdots a_{{\bf p}_n}(t)\; a^\dagger_{{\bf k}_1}\cdots a^\dagger_{{\bf k}_l}|0\rangle \\\\
        &= \frac{{}_t\langle0|0\rangle}{\sqrt{n!l!}}\left\{ \prod_{j=1}^n\left[ e^{-iE_{{\bf p}_j}t}f(t,{\bf p}_j) \sqrt{2E_{{\bf p}_j}} \right]\prod_{s=1}^l\left[ -f^*(t,{\bf k}_s) \sqrt{2E_{{\bf k}_s}} \right] \right.\\
         & + (2\pi)^3 \sum_{j=1}^n\sum_{s=1}^l (2E_{{\bf k}_s}) \, e^{-iE_{{\bf k}_s}t} \, \delta^3({\bf p}_j- {\bf k}_s)\prod_{j^\prime\neq j}\left[ e^{-iE_{{\bf p}_{j^\prime}}t}f(t,{\bf p}_{j^\prime}) \sqrt{2E_{{\bf p}_{j^\prime}}} \vphantom{\int }\right]\prod_{s^\prime\neq s}\left[ \vphantom{\int }-f^*(t,{\bf k}_{s^\prime}) \sqrt{2E_{{\bf k}_{s^\prime}}} \right]\\
         & \quad + \quad \left. \vphantom{\int }\mbox{ all combinations with two and more Dirac deltas} \quad\right\}.
    \end{split}
\end{equation}
The second equality is gained using the commutation relation \eqref{eq: commutation a(t) a(0)} and the actions on the vacuum \eqref{prop_to_coh}, it gives one term without any Dirac Delta, $nl$ (all the combinations between $p$'s and $k's$) with a single delta. If both $n$ and $l$ are larger than $1$, then there are $n(n-1)l(l-1)$ terms with two deltas (all the combinations for two contractions), and so on until either $n-n$ or $l-l$ is hit. The numerical factor is the same in all instances except that the functions for the indices that appear in the deltas are omitted.
\\
To finish the calculation, one has to compute the overlap ${}_t\langle0|0\rangle$. The modulus can be obtained by the normalization of the physical states, for instance $|0\rangle$:

\begin{equation}
    \begin{split}
    \label{overlap vac t and 0}
        1&=\langle 0 | 0 \rangle = \langle 0 | \, U^{\dagger}(t,0) \,\,U(t,0)\,| 0 \rangle = \sum_{n=0}^\infty \left[ \prod_{j=1}^n \int\frac{d^3 p_j}{(2\pi)^3 2 E_{{\bf p}_j}} \right] \left| \langle {\bf p}_1, \cdots, {\bf p}_n | \, U(t,0) \, |0\rangle \right|^2 \\
        &=  \sum_{n=0}^\infty \frac{1}{n!} \left[ \prod_{j=1}^n \int\frac{d^3 p_j}{(2\pi)^3} \right] \left| {}_t\langle 0|   \, a_{{\bf p}_1}(t)\cdots a_{{\bf p}_n}(t)|0\rangle \right|^2
        \\
        &=\sum_{n=1}^\infty\frac{\left|{}_t\langle 0|0\rangle \right|^2}{n!} \left( \int \frac{d^3 p}{(2\pi)^3} \; \left| f(t,{\bf p}) \right|^2 \right)^n = \left|{}_t\langle 0|0\rangle \right|^2 e^{N_0(t)} \text{ ,}
    \end{split}
\end{equation}
Where we recall $N_{0}(t)$ is the expectation value of the particle number operator given in \eqref{exp value 1}, leading to
\begin{equation}\label{mod}
    \left|{}_t\langle 0|0\rangle \right|= e^{-\frac{1}{2}N_0(t)} \text{ .}
\end{equation}

In Eq. \eqref{overlap vac t and 0} we recognize the probabilities for the vacuum to decay into $n$-particle states after a time $t$. The use of $t$ states simplifies the calculation of these to a few lines, compared to a more standard procedure of projecting onto eigenstates of the Hamiltonian. It is given by a Poissonian distribution with expectation value $N_{0}(t)$ (as expected), and the normalization of the probabilities is guaranteed by the normalization of the initial time vacuum state $\ket{0}$, i.e. \eqref{overlap vac t and 0}.  What remains is to calculate the phase of the overlap ${}_t\langle 0|0\rangle$. We leave the calculation of the phase to appendix \ref{sec:app_time_dep}. If the overlap is written as ${}_t\langle 0|0\rangle = \left|{}_t\langle 0|0\rangle \right| e^{i\varphi(t)}$, the phase $\varphi(t)$ is given by:
\begin{equation}
\label{phase phi}
\varphi(t) = \frac{i}{2} \int_0^t dx^0\int\frac{d^3 p}{(2\pi)^3} \left[ \partial_{x^0} f^*(x^0,{\bf p}) \, f(x^0,{\bf p}) - f^*(x^0,{\bf p}) \, \partial_{x^0} f(x^0,{\bf p}) \vphantom{\frac{}{}} \right] \text{ .}
\end{equation}

\subsection{One particle scattering}
We consider here the scattering of a single particle against the classical static potential generated by $g\rho({\bf x})$ and show in which sense we recover the intuitive notion of ``the quantum particle seems free as long as it misses the target''.

We take  a single particle initial state 

\begin{equation}
   |\psi_1\rangle = \int\frac{d^3 k}{(2\pi)^32 E_{\bf k}}\; \psi_1({\bf k})\; |{\bf k}\rangle,
\text{ .}\end{equation}
Moreover, we call $\psi_0(t;{\bf p}_1,\cdots ,{\bf p}_n)$ as the $n$-particle partial wave of the time evolving vacuum

\begin{equation}\label{psi_0_n 2}
    \psi_0(t;{\bf p}_1,\cdots ,{\bf p}_n) = \langle {\bf p}_1,\cdots ,{\bf p}_n| \, U(t,0) \, |0\rangle =  \frac{{}_t\langle0|0\rangle}{\sqrt{n!}} \prod_{j=1}^n\left[ e^{-iE_{{\bf p}_j}t}f(t,{\bf p}_j) \sqrt{2E_{{\bf p}_j}} \right],  
\end{equation}
which is computed from \eqref{scattering_time_dep_a}. We know its absolute value squared from \eqref{overlap vac t and 0}, but this intermediate result is convenient to recognize in the scattering amplitude formula. For $n = 0$ this is the survival amplitude of the particle vacuum which is given by
$  \psi_{0}(t) = {}_t\langle0|0\rangle$, see Eq. (\ref{mod}) for its modulus.
Only two terms in \eqref{scattering_time_dep_a} are nonzero in the one-particle scattering and they are given by:
\begin{equation}\label{scattering_1}
   \begin{split}
        \langle {\bf p}_1 \cdots {\bf p}_n | \, U(t,0)\, |\psi_1\rangle & = -\psi_0(t;{\bf p}_1 \cdots {\bf p}_n) \int\frac{d^3 k}{(2\pi)^3\sqrt{2E_{\bf k}}} \; f^*(t, {\bf k}) \,  \psi_1({\bf k}) \\
        & + \frac{1}{\sqrt n}\sum_{j=1}^n\left[ \psi_1({\bf p}_j) \, e^{-iE_{{\bf p}_j}t}  \; \psi_0(t;{\bf p}_1, \cdots{\bf p}_{j-1},{\bf p}_{j+1},\cdots, {\bf p}_n)\right].
   \end{split}
\end{equation}
The simplest cases to look at are for $n=0$ or $n=1$. In the $n=0$ case the only term is the absorption of the incoming particle by the source, but nothing is emitted as a result. For $n=1$ there is a contribution where the particle is absorbed and re-emitted, and a contribution where it does not interact with the source and propagates freely. Diagrammatically, for $n=1$ the amplitude can be understood as
\begin{equation}
\includegraphics[width=0.3\textwidth,valign=c]{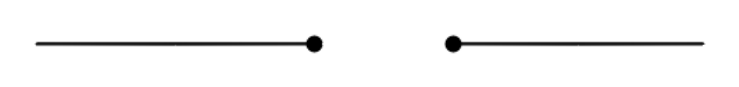}+\includegraphics[width=0.2\textwidth,valign=c]{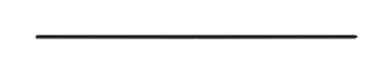},
\end{equation}
corresponding to the amplitude given by
\begin{equation}
\langle {\bf p} | \, U(t,0)\, |\psi_1\rangle  = -\psi_0(t;{\bf p}) \int\frac{d^3 k}{(2\pi)^3\sqrt{2E_{\bf k}}} \; f^*(t, {\bf k}) \,  \psi_1({\bf k}) \\
        + \left[ \psi_1({\bf p}) \, e^{-iE_{{\bf p}}t}  \; \psi_0(t)\right] \text{ ,}
\end{equation}
where: (i) the first term on the r.h.s.  contains the amplitude for the emission of a particle with momentum ${\bf p}$ out of the vacuum, $\psi_0(t;{\bf p})$ (that includes all vacuum contributions with creation and annihilation of an arbitrary number of particles) multiplied by the amplitude for the annihilation of the original state $|\psi_1\rangle$; (ii) the second term on the r.h.s. contains the amplitude for the free propagation of $|\psi_1\rangle$ multiplied by the vacuum amplitude $\psi_0(t)$, that includes the creation and subsequent annihilation of any number of particle.

While the pictures above are, strictly speaking, not Feynman diagrams, they offer an intuitive picture of the process. For general $n$ the structure is largely similar. In general the amplitude can be diagrammatically written as
\begin{equation}
\includegraphics[width=0.3\textwidth,valign=c]{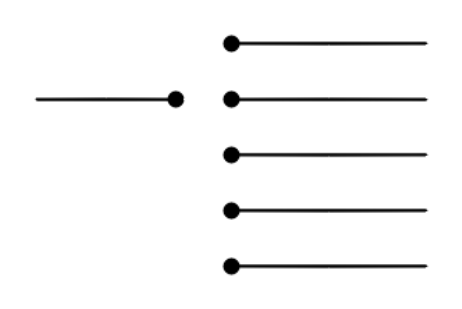}+\includegraphics[width=0.3\textwidth,valign=c]{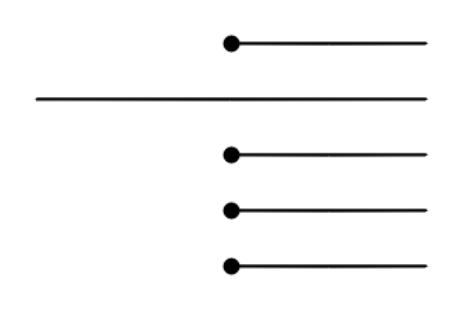}\, .
\end{equation}%
The first term can be understood as the probability amplitude that the starting particle gets absorbed, times the $n$-particle wave-function of the vacuum. The second part is the free-particle wave-function $\psi_i({\bf p})e^{-i E_{\bf p}t}$ times the $(n-1)$-particles vacuum contribution, where the $\mathbf{p}_{j}$ is not in the vacuum contribution. The sum guarantees that this contribution is manifestly symmetric in the exchange of any particle, and the $1/{\sqrt n}$ makes sure it is normalized to $1$, as is mandatory for physical states. The entire dynamics of the system can be summed up as an interference pattern between the decaying vacuum, a probability amplitude of absorbing the original particle and a free-like evolution of the original state. The vacuum decay is a key component of the scattering; if the vacuum did not decay, there would have been no scattering. 
\\
Focusing on the absorption amplitude, calling  $\psi_1(x)$

\begin{equation}
    \psi_1(x) = \int \frac{d^3 p}{(2\pi)^32E_{\bf p}} \; e^{-ip\cdot x} \psi_1({\bf p}),
\end{equation}
the wave function in space-time that the original state would have if it were a free system, making use of the definition of $f(t,{\bf k})$ in~(\ref{g_f}), one has

\begin{equation}
    \begin{split}
        {\cal A}(t)&=-\int\frac{d^3 k}{(2\pi)^3\sqrt{2E_{\bf k}}} \; f^*(t, {\bf k}) \,  \psi_1({\bf k}) = -i\, g \int_0^tdy^0\int d^3y \int\frac{d^3 k}{(2\pi)^32E_{\bf k}} \;\rho({\bf y}) \,  e^{-ip\cdot y}\psi_1({\bf k}) \\
        &=-i \int_0^tdy^0\int d^3y \; g \,  \rho({\bf y}) \;  \psi_1(y).
    \end{split}
\end{equation}
The absorption amplitude can also be seen as a functional ${\cal A}[g\rho, \psi_1]$. This amplitude vanishes if the time integral of the overlap between the free-like wave-function $\psi_1(x)$ and the classical source is vanishing. The absorption amplitude could be zero for another reason, for example if $\psi(x)$ is an even function of space and $\rho(\mathbf{x})$ is an odd function of space. The simplest case is, of course, that the support of $\rho$ and $\psi_1$ is mutually exclusive during the whole time of the evolution

\begin{equation}
    \rho({\bf x}) \psi_1(x)=0, \qquad \forall \, x^0\in[0,t], \; \forall\, {\bf x}\in {\mathbb R}.
\end{equation}
This corresponds to the intuitive notion of ``the wave-packet misses the classical source''. Because of the exact and general~\eqref{scattering_1}, if the amplitude ${\cal A}$ is vanishing the only remaining part of the dynamics is the eventual decay of the vacuum, in the form of the $\psi_0$'s, with an appropriate totally symmetric combination of the free-like part and the vacuum contribution. It is not possible to remove this vacuum decay, but we can look at the conditional probability that no extra particles come from the vacuum decay. That is, we normalize the probability amplitudes by dividing by $\int \frac{d^3\mathbf{p}}{(2\pi)^3 2E_{\mathbf{p}}}|\langle{\bf p}|U|\psi_1\rangle|^2$, so the probabilities are conditional on the observation of `one particle goes in, one particle comes out', while keeping the original phase in the wave functions. (In turn, this operation also corresponds to the `collapse' of the wave function if the final state is measured to contain one particle only). 
By construction, the resulting wave functions are normalized to 1. When the absorption amplitude is zero, the resulting one particle wave function after scattering is
\begin{equation}
    e^{i\varphi(t)} \; e^{-iE_{\bf p}t} \psi_1(\bf p) \text{ .}
\end{equation}
This is `almost' the wave-function of a free particle in the Shr\"odinger representation. 
In fact, the energy phase $e^{-iE t}$ is responsible of the time evolution of the states, even for free fields. The phase $e^{i\varphi}$ is not, but it can be removed from the theory without changing the Euler-Lagrange equations of motion since it corresponds to a (time-dependent) phase-convention of the basis, rather than any physical observable. 
\\
Since the $\varphi(t)$ itself is given as a time integral ~\eqref{phase phi}, it is possible to add to the Lagrangian of the system the function $-\partial_t\varphi(t)$ to compensate exactly in the final formulas.
\\
In this sense then, for a vanishing overlap ${\cal A}(t)=0$, the elastic scattering with the classical source $g\rho$ is trivial and indistinguishable from a free particle. The non-trivial dynamics involve the creation of and interference with the eventual extra particles from the vacuum decay. 
\\
It goes without saying that the elastic scattering is not trivial for a non-vanishing ${\cal A}$. In that case the normalized partial wave is given by:

\begin{equation}
    \left[ {\cal A}(t) \, \psi_0(t;{\bf p}) \; + \;  \psi_0(t) \, e^{-iE_{\bf p}t} \psi_i({\bf p}) \vphantom{\int} \right]\frac{\exp[\frac{1}{2}N_{0}(t)]}{\sqrt{|{\cal A}(t)|^2\left[  \vphantom{\frac{}{}}N_0(t)-2\right]+1}},
\end{equation}
or just $0$ if $\sqrt{2E_{\bf p}}f(t,{\bf p})$ happens to be, at time $t$, exactly equal to the starting $\psi_1$, $f(t,{\bf p})\sqrt{2E_{\bf p}} = \psi_1(\bf p)$.
\\
All of this can be extended to the case of many particles initial states. Instead of one absorption amplitude there are amplitudes for the absorption of one or more of the starting particles, the vacuum decays as in the one particle case, and (parts of \footnote{In the case of some of the particles being absorbed, but not all off them, only the remaining contribution of the initial state evolves just with the phases $e^{-iE_{{\bf p}_j}t}$. There is not only the case of no-particle absorbed, and a free evolution of the initial state. }) the initial state that evolves freely.

\section{Vacuum decay and Schwinger mechanism}
\label{sect:vacuum decay}
In the previous section we have seen that the initial particle vacuum $\ket{0}$ is not the lowest energy eigenstate (it is not even an energy eigenstate) and so it will decay and produce particles. We have a situation conceptually quite similar to the Schwinger mechanism, but there are some interesting differences in the result. In \eqref{overlap vac t and 0} we can see the probability distribution for the particle vacuum to decay into $n$ particle after time $t$ has elapsed. It is a Poisson distribution with the expectation value found earlier in \eqref{exp value 1}, that is
\begin{equation}
\begin{split}
    \label{expectation value}
&P_{0\to n}(t) = \frac{e^{-N_0(t)}}{n!}\left[ \vphantom{\frac{}{}} N_0(t) \right]^n,
\\    
     &N_0(t) = \bra{0}\hat N(t)\ket{0} =  2\int \frac{d^3p}{(2\pi)^3} \left( \vphantom{\frac{}{}} 1- {\rm cos}(E_{\bf p}t)  \right) \left|  h({\bf p})\right|^2.
     \end{split}
\end{equation}
The function $h(\mathbf{p})$ is a weighted Fourier transform of the source $g\rho(\mathbf{x})$, given in \eqref{g_f}.  As expected, in the limit of $t\to 0$ the expectation value is 0 and so the vacuum is stable. The limit $t \to \infty$ depends on the form of the source, but we can deduce it is finite. The integral can be re-written as an integral over energy and angle (which is bounded), and is absolutely convergent. Therefore the Riemann-Lebesgue lemma applies which says that in the limit of $t \to \infty$ the oscillatory term goes to 0, meaning the expectation value goes to the finite nonzero positive value:
\begin{equation}\label{eq: expectation value limit}
  \lim_{t \to \infty}  N_0(t) =  2\int \frac{d^3p}{(2\pi)^3}  \left|  h({\bf p})\right|^2 <\infty \text{ ,}
  \end{equation}
implying that the particle production rate goes to zero for large times, leaving us with a finite amount of particles (or particle density for large volumes), without any need to turn off the source. Since the derivation of section \ref{Decay probability} was valid even for time dependent sources, the source can of course be chosen so that it turns off after some time.

{It is interesting to note that there is no state that fulfills a generalized definition of a vacuum that has zero particle number expectation value for all times. This can be seen by looking at time $t=0$, for which an arbitrary pure state $\ket{\psi}$ has particle number expectation value}
\begin{equation}
        \int \frac{d^3 p}{(2\pi)^3} \bra{\psi} a_{\mathbf{p}}^{\dagger} a_{\mathbf{p}}\ket{\psi} = \int \frac{d^3 p}{(2\pi)^3} \left| a_{\mathbf{p}} \ket{\psi}  \right|^2.
\end{equation}
{This is only zero when $a_{\mathbf{p}} \ket{\psi} = 0$ for all $\mathbf{p}$. Having no quantum degree of freedom other than the field $\hat \phi$ (no other fields, scalar or otherwise), this is only true for the state $\ket{0}$. This argument generalizes to mixed states as well. Since $\ket{0}$ is the only state that fulfills zero particle number at $t=0$, and for $t>0$ it has a nonzero particle number expectation value as per \eqref{exp value 1}, we conclude a state with zero particle number expectation value for all time does not exist.}

\subsection{Quantum Zeno effect}
For short times there is a quadratic behavior that allows for the quantum Zeno effect. Namely, the vacuum survival probability 
\begin{equation}
\label{vacuum survival probability}
     P_{0 \to 0}(t)={\rm exp}\left\{-2\int\frac{d^3q}{(2\pi)^3}\left(1- {\rm cos} (E_{\bf q}t) \vphantom{\frac{}{}}\right)\left|  h({\bf q})\right|^2 \right\}
\end{equation}
can be expanded for short times:
\begin{equation}
    {\rm exp}\left\{-\int\frac{d^3q}{(2\pi)^3} E_{\bf q}^2t^2 \vphantom{\frac{}{}}\left|  h({\bf q})\right|^2 \right\} = e^{- \frac{ t^2 }{\tau_{Z}^2} } \text{ ,} 
    \label{eq:small time behaviour}
\end{equation}
where we define a `Zeno time' $\tau_{Z}$ by 
\begin{equation}
    \frac{1}{\tau_{Z}^2}\equiv \int\frac{d^3q}{(2\pi)^3} E_{\bf q}^2 \vphantom{\frac{}{}}\left|  h({\bf q})\right|^2.
\end{equation}
We may look at the situation where we have repeated measurements of the vacuum and whether or not it survived. We let the system evolve over a large time $T$ with $N$ measurements separated by equally spaced (small) time intervals $t = T/N$. For the vacuum to survive at time $T$, it needs to survive at each measurement, meaning
\begin{equation}
    P_{0 \to 0}(T) = (P_{0 \to 0}(t))^{N} \approx \exp\left[-N \frac{t^2}{\tau_{Z}^2} \right] = \exp \left[-\frac{1}{N}\frac{T^2}{\tau_{Z}^2}\right].
\end{equation}
The small time behaviour of the survival probability now also holds for arbitrarily long times $T$, given $N$ is chosen sufficiently large so that $t$ is sufficiently small. Furthermore, for a given time $T$, $N$ can be chosen arbitrarily large, which makes the survival probability approach 1. This is an explicit example of the quantum Zeno effect \cite{Misra:1976by,Fonda:1978dk,Itano:1990zz,Koshino:2004rw,Giacosa:2011xa}: by repeatedly measuring the system we can preserve the initial state which would normally be prone to change over time. It is a consequence of a decay that is slower than an exponential decay for small times. The quantum Zeno effect is a generic feature of theories obeying quantum mechanics. It is found for example in cavity QED \cite{Haroche:2014gla,2008PhRvL.101r0402B} and may even affect Hawking radiation in large black holes \cite{Nikolic:2013pza}. Moreover, upon inspection of \eqref{vacuum survival probability}, we can reason there is no regime where an exponential decay law is a good approximation. For small times it is given by \eqref{eq:small time behaviour}, after which the oscillatory behaviour becomes relevant. At large time the oscillation dies out and the particle number expectation value becomes a non-vanishing constant.
\subsection{Numerical examples}
\label{sect:numeric}
\begin{figure}
    \centering
    \includegraphics[width=\textwidth]{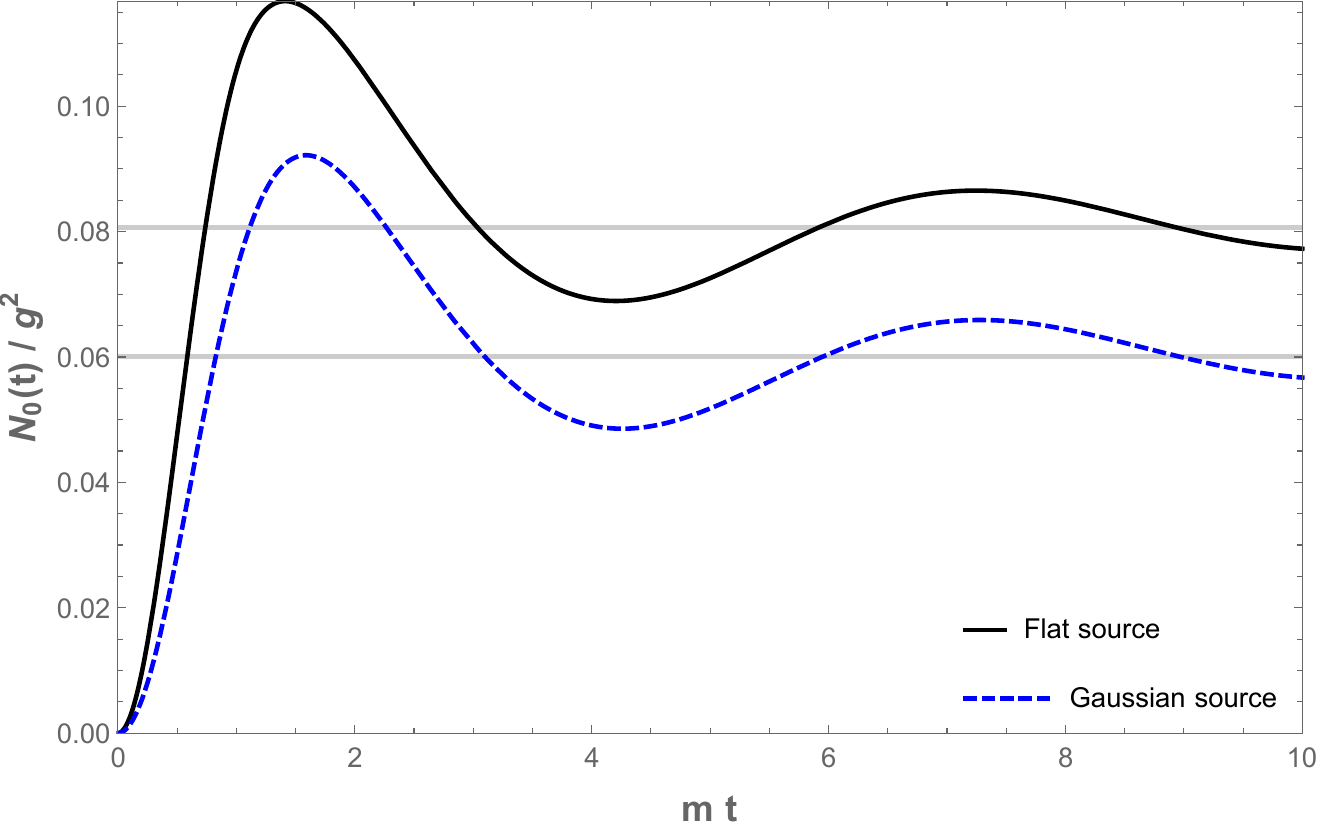}
    \caption{Expectation value for the number of particles produced for the different sources as a function of time. The parameters $g$ and $m$ are absorbed in the expectation value and time units respectively, and we use $L = 1$ (in $1/m$ units). The grey lines are the infinite time asymptotes found by setting the oscillatory term to 0.}
    \label{fig:expectation1}
\end{figure}

\begin{figure}
    \centering
    \includegraphics[width=\textwidth]{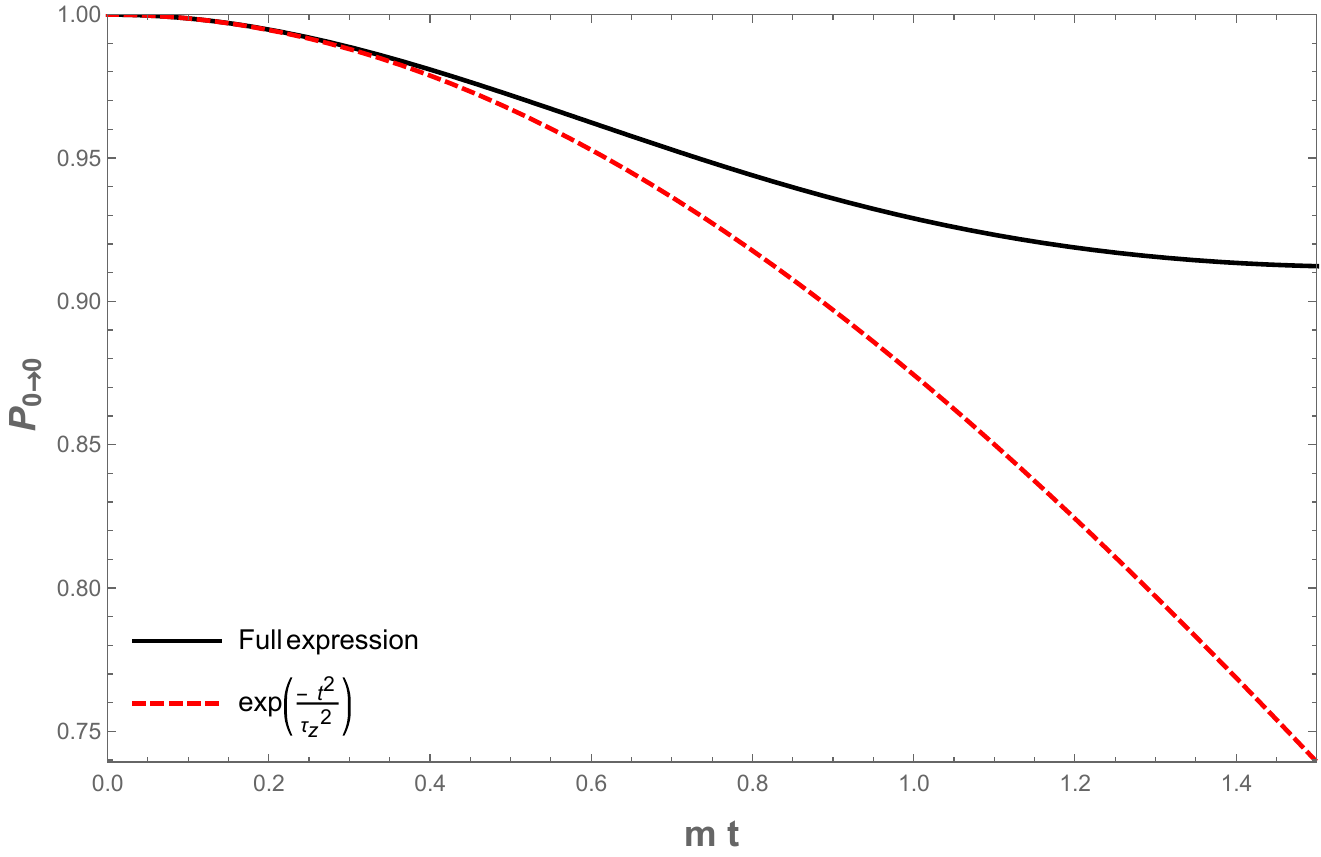}
    \caption{Vacuum survival probability with a Gaussian source, compared with the small time behaviour found in \eqref{eq:small time behaviour}. The parameter  $m$ is absorbed in the time units, and $g=1$ and $L = 1 $ are used.}
    \label{fig:zeno gaussian}
\end{figure}

\begin{figure}
    \centering
    \includegraphics[width=\textwidth]{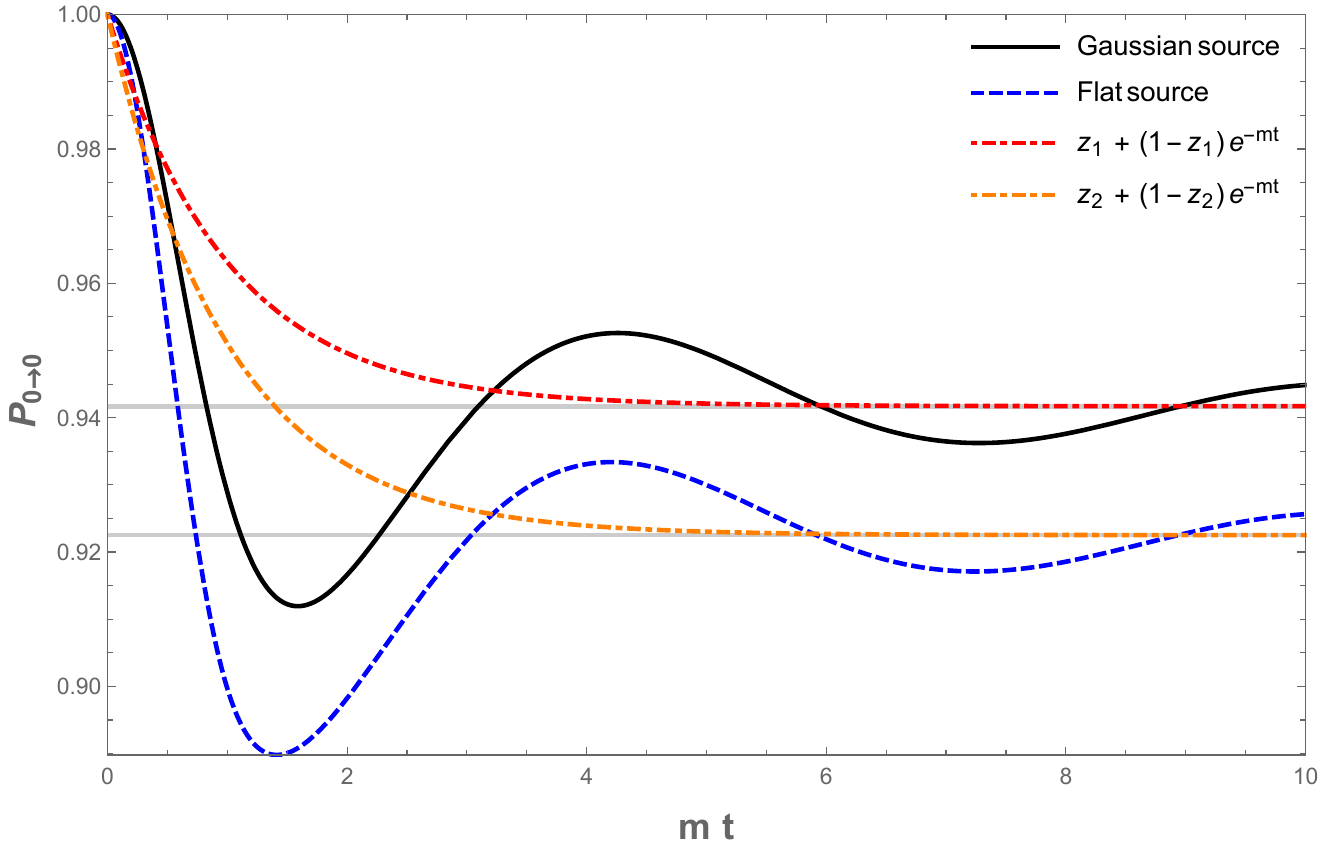}
    \caption{Vacuum survival probability with both sources with a longer timescale, compared with a exponential decay which saturates to a non-zero value at infinity. The parameter $m$ is absorbed in the time units, and we set $ L = 1$ in units of $1/m$, and $g=1$. The grey lines indicate the asymptotic values for the survival probability for the two different sources.}
    \label{fig:exponential approx}
\end{figure}

\noindent
As a computational example, and to illustrate the time dependent behaviour, we shall compute the probability distribution for two different forms of the external source, a Gaussian and a flat source. We will take them of the forms:
\begin{align}
   g \rho_1({\bf x}) &= g\, m^3 \,  \exp \left(-\,\pi \frac{\bf x^{2}}{L^{2}}\right), \\
   g \rho_{2}(\mathbf{x}) &= g\, m^3\, \theta \left(\frac{L}2-x\right) \theta \left(\frac{L}2+x\right) \theta \left(\frac{L}2-y\right) \theta \left(\frac{L}2+y\right) \theta
   \left(\frac{L}2-z\right) \theta \left(\frac{L}2+z\right) \text{ ,}
\end{align}
where $g$ plays the role of a dimensionless coupling constant determining the strength of the interaction, and $L$ is a parameter that determines the volume of the sources. The sources are normalized such that their peaks and total charges are equal $\int d^{3}\mathbf{x} g\rho_{1}(\mathbf{x}) = \int d^{3}\mathbf{x} g\rho_{2}(\mathbf{x}) = g \, m^3 L^{3} $, and $\rho_{1}(0)=\rho_{2}(0) = g\, m^3$. For the expectation value we need the weighted Fourier transforms of the sources (see Eq. (\ref{g_f})), which are given by
\begin{align}
  h_{1}(\textbf p) &= - \frac{g}{(\mathbf{p^2}+m^2)^{3/4}}   \, m^3 L^{3} \, e^{-\frac{ p^2 L ^2}{4\pi}}, \label{h function gaussian}
  \\
  h_{2}(\mathbf{p}) &= - \frac{g}{(\mathbf{p^2}+m^2)^{3/4}} \frac{8 m^3 \sin \left(\frac{L p_{x}}2\right)\sin \left(\frac{L p_{y}}2\right)\sin \left(\frac{L p_{z}}2\right)}{p_{x}p_{y}p_{z}},\label{h function flat}
\end{align}
where $p_x, p_y, p_z $ are the components of the 3-momentum $\mathbf{p}$. This allows us to compute the integral \eqref{expectation value} numerically. Since the mass is the only parameter with dimension of energy for this form of the sources, we will set the mass $m=1$. Consequently the time $t$ and length-scale $L$ will be in units $1/m$. The dependence on $g$ is simple: $N_{0}(t)$ scales with $g^2$ (even though the result is non-perturbative), and the dependence on $L$ has no simple form except in the large L limit, which we look at shortly. The result of this with the parameters $g$ and $L$ set to 1 is shown in figure \ref{fig:expectation1}, where  we can see that saturation of the particle expectation value sets in quite quickly. The vacuum survival probability for the Gaussian source, together with an exponential decay and the small time behaviour are shown in figure \ref{fig:zeno gaussian}. 
\\
Taking, as an illustrative example, the case of the dilaton \cite{Migdal:1982jp}, assuming a mass of about 1.7 GeV, and using $\hbar = 6.6 \cdot 10^{-25}$ GeV s = 1, units of time are of approximately $4 \cdot 10^{-25}$ seconds. 
From figure \ref{fig:expectation1} the particle content is close to saturation after 10 units of $m\,t$ (or earlier), a typical timescale for the dilaton in this case is in the order of $10^{-24}$ seconds, or $\sim 1\, \text{fm}/c$.
\\
In figure \ref{fig:exponential approx} a comparison of the vacuum survival probability with an exponential approximation of the form $z_{i} + (1-z_{i})\exp{-\Gamma t}$, where $z_{i}$ is the saturation value found by exponentiating \eqref{eq: expectation value limit}, the index $i = 1,2$ denotes to which shape of the source i.e. \eqref{h function gaussian} or \eqref{h function flat}, it belongs to, and where we have taken $\Gamma \sim m$. We have also taken $g=1$, but scaling to $g \neq 1$ is done simply by taking the probability to the power of $g^2$. Note, the choice $\Gamma = m$ here is arbitrary and is chosen just to `guide the eye' in order to manifestly show that it does not lead to a good description of the actual survival probability. The only regime where the exponential approximation is close is at large times when the fluctuations around the saturation value are too small. A better description of the theoretical curve could be obtained  by an exponential with superimposed oscillations as well as an offset, but we leave this point for future phenomenological investigations. 

It is interesting to observe that similar oscillations (but not decays) arise in the case of mixing of bosonic  fields \cite{Blasone:2001du,Blasone:2023brf}, which in turn might be relevant for neutrino oscillations \cite{Smaldone:2021mii}.


\subsection{Large volume limit}
\noindent
The scenario where the size of the classical source is much larger than the scales probed by the system is of course a simplification which is however interesting for practical purposes. In our system, this large volume limit can be explicitly computed for both sources. Starting with the Gaussian shaped source, the expectation value per volume element $L^3$ is found using \eqref{expectation value}:
\begin{align}
\label{eq:gaussian volume derivation}
      \frac{1}{L^3} N_0(t) = 2 \int \frac{d^3\mathbf{p}}{(2\pi)^3}\left(1-\cos(E_{\mathbf{p}}t)\right) \frac{g^2}{(\mathbf{p}^2+m^2)^{3/2}} m^6 L^{3} \, e^{-\frac{ p^2 L ^2}{2\pi}}.
\end{align}
For taking the large volume limit, that is $L \to \infty$, we will want to identify representations of the Dirac delta function. In this case for the Gaussian source, the relevant representation of the Dirac delta is
\begin{equation}
   \delta(x) = \lim_{\epsilon \to 0^{+}} \frac{e^{-x^2/4\epsilon}}{2\sqrt{\pi\epsilon}} 
\end{equation}
Renaming $\frac{L^2}{2\pi} = \frac{1}{4\epsilon}$ on the RHS of \eqref{eq:gaussian volume derivation} this form of the Dirac deltas will be more easily identifiable:
\begin{align}
    \frac{1}{L^3} N_0(t) = 2 \int \frac{d^3\mathbf{p}}{(2\pi)^3}\left(1-\cos(E_{\mathbf{p}}t)\right) \frac{g^2 m^6}{(\mathbf{p}^2+m^2)^{3/2}} \pi^3 \sqrt{8} \left(\frac{e^{-p_{x}^2/4\epsilon}}{2\sqrt{\pi\epsilon}}\right)\left(\frac{e^{-p_{y}^2/4\epsilon}}{2\sqrt{\pi\epsilon}}\right)\left(\frac{e^{-p_{z}^2/4\epsilon}}{2\sqrt{\pi\epsilon}}\right).
\end{align}
When we take the large volume limit $L \to \infty$ which implies $\epsilon \to 0^{+}$ the last 3 factors approach delta functions of the components of $\mathbf{p}$. Evaluating the momentum integral with these delta functions gives
\begin{align}
     \frac{1}{L^3} N_0(t) = \frac{g^2}{\sqrt{2}} m^3(1-\cos(mt)).
     \label{eq:gaussian volume limit1}
\end{align}
Similarly, in the computation for the square source we will need the following representation of the Dirac delta:
\begin{equation}
    \delta(x) = \lim_{L \to \infty} \frac{\sin^{2}(L \, x)}{\pi L  \,x^2}
\end{equation}
Using the weighted Fourier transform of the flat source \eqref{h function flat} in the expectation value \eqref{expectation value} gives
\begin{align}
    \frac{1}{L^3} N_0(t) = 2 \int \frac{d^3\mathbf{p}}{(2\pi)^3}\left(1-\cos(E_{\mathbf{p}}t)\right) \frac{g^2}{(\mathbf{p}^2+m^2)^{3/2}}  \frac{64 m^6}{L^3}\left[\frac{\sin^2(Lp_{x}/2)}{p_x^2}\right] \left[\frac{\sin^2(Lp_{y}/2)}{p_y^2}\right] \left[\frac{\sin^2(Lp_{z}/2)}{p_z^2}\right].
\end{align}
Once again, we bring it into the form for Dirac deltas:
\begin{align}
    \frac{1}{L^3} N_0(t) = 2 \int \frac{d^3\mathbf{p}}{(2\pi)^3}\left(1-\cos(E_{\mathbf{p}}t)\right) \frac{g^2 m^6}{(\mathbf{p}^2+m^2)^{3/2}} \pi^3 \left[\frac{\sin^2(Lp_{x}/2)}{\pi L (p_x/2)^2}\right] \left[\frac{\sin^2(Lp_{y}/2)}{\pi L (p_y/2)^2}\right] \left[\frac{\sin^2(Lp_{z}/2)}{\pi L (p_z/2)^2}\right]
\end{align}
In the limit $L \to \infty$ the sinc functions become delta functions of the components of $\mathbf{p}/2$ which equal $8$ times the delta function of $\mathbf{p}$ and we find
\begin{align}
    \frac{1}{L^3} N_0(t) = 2g^2 m^3(1-\cos(mt)).
\end{align}
Note that taking this limit does not commute with taking the $t \to \infty$ limit, as here the oscillation does not die down anymore since the integral is already evaluated. Taking the infinite time limit before the volume limit, the oscillatory term vanishes while leaving the rest of the derivation entirely the same. Then the large volume and infinite time limit is
\begin{equation}
\begin{split}
    \lim_{L \to \infty} \lim_{t \to \infty} \frac{1}{L^3}N_0(t) &= 2g^2 m^3 \qquad \phantom{\frac{1}{\sqrt{2}}}(\text{flat source}),
    \\
    \lim_{L \to \infty} \lim_{t \to \infty} \frac{1}{L^3}N_0(t) &= \frac{g^2}{\sqrt{2}}m^3 \qquad \phantom{2} (\text{Gaussian source}).
    \end{split}
\end{equation}
Most of the form can be found just using dimensional arguments, since both sides are dimensionless and the dependence on $g$ is just $g^2$, but the prefactor is dependent on the shape of the source and on how the volume is defined. In this sense the flat source gives the same result using the informal statement `$\delta^{3}(0) \to (2\pi)^3 V$', but for the Gaussian source it does not.

These results are differs from the Schwinger mechanism, where the particle production rate is constant, so the number of particles is linear in time and diverges for $t \to \infty$. This divergence is of course unphysical, and it is tamed by the adiabatic hypothesis which turns off the source for large times. 
In our situation the particle production rate goes to zero for large times, leaving us with a finite amount of particles (or particle density for large volumes), without any need to turn off the source. Since the derivation of section \ref{Decay probability} was valid even for time dependent sources, the source can of course be turned off, even in a non-adiabatic way.


\section{Discussion}
\label{comparison to literature}

The quantum field theory of a scalar field coupled to some external source is one of the simplest QFTs one can write down, it has been considered in the past and it is not surprising that it has been mentioned or discussed in textbooks, see e.g. \cite{Peskin:1995ev,Coleman:2018mew,Wentzel:1949, Ticciati:1999qp}. It is commonly accepted that the dynamics can be non-trivial for a classical source $\rho(t,{\bf x})$ which is Fourier transformable in spacetime, hence time dependent. The case of a time independent source is treated as a limiting case of the former. In particular, the originally time-independent classical source is multiplied by an adiabatic function that switches off the interaction in the long past and future. The scattering matrix in the infinite time limit of persistence of the interaction is essentially the identity matrix, and this is usually assumed to imply an equivalence between the interacting system (without the adiabatic function) and a free quantum field. Since some of the intuitive statements commonly used are incomplete, and it is easy to miss some fundamental conceptual steps looking just at the scattering matrix, it is convenient to have a more detailed discussion.

This section is dedicated to the paradoxical comparison between the results of this paper, in which the dynamics is not trivial, and the past literature. We will refer especially to the seminal lectures on QFT by Coleman~\cite{Coleman:2018mew}, since the treatment is rather extensive and it inspired some of the other books, e.g. \cite{Ticciati:1999qp}, especially regarding the triviality of a time independent source. The starting point is the exact solution for the time dependent case, obtained resumming all the terms of the Dyson series~\cite{Coleman:2018mew}. Using the notation in this work, the scattering matrix reads

\begin{equation}\label{S_time_dep}
\begin{split}
    S &= U_I(\infty,-\infty) = :{\rm exp}\left\{ -\frac{g^2}{2}\int d^4 x\int d^4 y \, \rho(x)\rho(y) \, \Delta_F(x-y) \right. \\
    & \left. - i g \int d^4 x \int\frac{d^3 p}{(2\pi)^3\sqrt{2 E_{\bf p}}}\left[ \rho(x) e^{-ip\cdot x} \, a_{\bf p} + \rho(x) e^{ip\cdot x} \, a^\dagger_{\bf p} \vphantom{\frac{}{}}\right]\right\}: \, .
\end{split}
\end{equation}
This formula makes sense for a fast decreasing (and hence Fourier transformable in spacetime) classical source $\rho(x)$. In particular the field is asymptotically free in the far past and far future $(\Box +m^2)\hat\phi\to 0$ as $t\to\pm\infty$ by the fast decreasing (in time too) assumption on the source. Making use of the energy integrated form of the Feynman propagator

\begin{equation}
    \Delta_F(x-y) = \int \frac{d^3 p}{(2\pi)^32 E_{\bf p}} \left[ \vphantom{\frac{}{}} \theta(x^0-y^0)e^{-ip\cdot(x-y)} + \theta(y^0-x^0)e^{ip\cdot(x-y)} \right],
\end{equation}
it is rather straightforward\footnote{The fastest way is to notice that the relations~(\ref{prop_to_coh}) between the basis $|\cdots\rangle_t$ and $|\cdots\rangle$ at $t=0$ can be generalized to the case $|\cdots\rangle_t\leftrightarrow|\cdots\rangle_{t_0}$ using the very same arguments as before, the difference being the function $f$ is integrated from $t_0$ in the lower bound instead of $0$ as in~\eqref{gen_f}. Therefore the formulas in~(\ref{t_states}) generalize to the $t_0$ basis using $U(t,t_0)$ instead of $U(t,0)$ as it could have been expected.} to show that the first term in the right hand side of~(\ref{S_time_dep}) is just ${}_t\langle 0|0\rangle_{t_0}$ in the $t\to \infty$ and $t_0\to -\infty$ limit. The term in the creation operator reads
\begin{equation}\label{operatorial_S_cr}
    {\rm exp}\left\{ -\int \frac{d^3p}{(2\pi)^3} \, f(\infty,-\infty;E_{\bf p},{\bf p}) \, a^\dagger_{\bf p} \vphantom{\frac{}{}} \right\},
\end{equation}
and the analogous for the term in the destruction operators is
\begin{equation}\label{operatorial_S_de}
    {\rm exp}\left\{ \int \frac{d^3p}{(2\pi)^3} \, f^*(\infty,-\infty;E_{\bf p},{\bf p}) \, a_{\bf p} \vphantom{\frac{}{}} \right\} \text{ .}
\end{equation}
 
Because of the normal ordered product the two terms factorize, with the creation term on the left and the destruction one on the right\footnote{Within normal ordered products the creation and destruction operators commute. This is the usual case of factorization of the exponential of the sum of two commuting operators. The normal ordering symbol can be removed as long as the exponent on the creation operators is on the right.}. The notation used for the $f$ function is the general (that admits time dependent sources) one as in~\eqref{gen_f} in Appendix~\ref{sec:app_ham}. The one used in the main text~\eqref{g_f} is thus a specific case. One can verify that the matrix elements of the scattering matrix~(\ref{S_time_dep}) coincide with the limit

\begin{equation}\label{scattering_time_limit}
   \langle {\bf p}_1,\cdots,{\bf p}_n|S|{\bf q}_1, \cdots, {\bf q}_m\rangle =  \lim_{{t_-\to-\infty}} \left[  \lim_{{t_+\to\infty}} \left(\vphantom{\frac{}{}} {}_{t_+}\langle {\bf p}_1,\cdots,{\bf p}_n|{\bf q}_1, \cdots, {\bf q}_m\rangle_{t_-} \right)\right],
\end{equation}
using the algebra of the time-dependent bases of Section~\ref{Decay probability}. The usual approach to treat the time-independent case is to assume that the scattering matrix can be obtained from~(\ref{S_time_dep}). Since the expression in~(\ref{S_time_dep}) requires a classical source which is Fourier transformable in time, and not only in the space coordinates, one considers the particular case of $\rho(t,{\bf x})= s(t)\rho(\bf x)$. The (smooth) adiabatic function $s(t)$

\begin{equation}\label{adiabatic_0}
    0<s(t)<1, \qquad s(t)=1 \, \forall |t|<\frac{T}{2}, \qquad \lim_{t\to\pm\infty}s(t)t^n=0 \quad \forall n\in {\mathbb Z}^+,
\end{equation}
effectively switches on(off) the interaction in the far past(future) in a Fourier transformable way and keeping the interaction the wanted (time-independent) one in the bulk $t\in\{-T/2,T/2\}$. The intuitive idea is that in the limit of $T\to\infty$ one has to recover the full time-independent theory, after all, in other cases, this procedure works. We show in this section that this is not the case, despite the fact that for a sudden switch $s(t)= \theta(t+T/2)\theta(T/2-t)$ equation~(\ref{scattering_time_limit}) is fulfilled in the $T\to\infty$ limit\footnote{Technically, both in the left hand side and in the righ hand side there is an undefined phase that diverges. It can be treated adding a constant in the Lagrangian (not the Lagrangian density) to off-set the minimum of the Hamiltonian and have a well-defined phase. This is completely analogous to the treatment in~\cite{Coleman:2018mew} of the phase in ``model II''.}. Already in the lectures~\cite{Coleman:2018mew} it is recognized that the conditions in~(\ref{adiabatic_0}) should be supplemented. Indeed, in order to recover the correct imaginary part of the function in~(\ref{S_time_dep}) in the $T\to \infty$ limit, it is necessary to add
\begin{equation}\label{adiabatic_1}
    s(t)= 0 \quad \forall |t|\ge T/2 +\Delta(T), \qquad \Delta(T)\ge 0, \qquad \lim_{T\to \infty} \frac{\Delta(T)}{T}=0 \text{ .}
\end{equation}
Later on, we shall simply write $\Delta$ for $\Delta(T)$.
In this way the Fourier transform of the adiabatic function fulfills the following relations

\begin{equation}\label{deltas}
    \int dt \, e^{\pm i E t} s(t) \xrightarrow{T\to \infty} (2\pi)\,\delta(E),\quad \frac{1}{T}\left| \int dt \, e^{\pm i E t} s(t) \right|^2\xrightarrow{T\to\infty}(2\pi)\,\delta(E).
\end{equation}
The conditions~(\ref{adiabatic_1}) are not really necessary for the first limit. Any adiabatic function would suffice. As it can be seen with a direct calculation\footnote{The function $s$ must be suppressed fast enough for $|t|>T/2$ to be able to use minorants of the modulus square outside of $t\in\{-T/2, T/2\}$ to show that they do not contribute. The rest of the proof consists just of the properties of the sinc delta family $[{\rm sin}(x/\epsilon)]/(\pi x)\xrightarrow{\epsilon\to 0}\delta(x)$ and the $\epsilon[{\rm sin}^2(x/\epsilon)]/(\pi x)\xrightarrow{\epsilon\to 0}\delta(x)$ one.}, it is needed for the second one.

At this point in Ref \cite{Coleman:2018mew} the operatorial parts~(\ref{operatorial_S_cr}) and~(\ref{operatorial_S_de}) are dismissed with an intuitive but deeply misleading argument. Since the function $f$ in~(\ref{operatorial_S_cr}) reads, in the presence of the adiabatic function

\begin{equation}
    f(\infty,-\infty;E_{\bf p},{\bf p})= \int dt \; e^{iE_{\bf p}t}\,s(t) \;\left[ i \frac{g}{\sqrt{2 E_{\bf p}}}\int d^3x \, \rho({\bf x}) \, e^{-i{\bf p}\cdot{\bf x}} \right],
\end{equation}
thus it approaches a Dirac delta as $T\to \infty$ according to the first equation in (\ref{deltas}). It is assumed that it goes to zero in this limit, the support of the Dirac Delta is just the origin, and the on-shell energy $E_{\bf p}\neq 0$ is never in the support for a massive field after all. This argument would be valid if $a^\dagger_{\bf p}$ were a test function rather than an operator. In the latter case the convergence properties in the space of generalized functions or distributions are irrelevant. In the lectures~\cite{Coleman:2018mew} it is presented an incorrect argument to argue that in the function space the integral of the adiabatic function is going to be vanishing for on-shell energies. It is based on the common misconception that, since some successions of functions converging to a Dirac delta are point wise vanishing outside of $0$ and sharply peaked in $0$, for instance a succession of Gaussians with a decreasing width $\sigma\to 0$, this must happen for any succession. This is clearly not the case for the family

\begin{equation}
    \delta_\varepsilon(x) = \sqrt{\frac{2}{\pi}} \, \frac{1}{\varepsilon} \, {\rm sin}\left( \frac{x^2}{\varepsilon^2} \right)\xrightarrow{\varepsilon\to 0} \delta(x),
\end{equation}
which is converging to a Dirac delta\footnote{This can be seen from the general construction of a family of regular functions converging to a delta. Any function $g(x)$ normalized to $1=\int d x\,  g(x)$, that doesn't spoil the uniform convergence of the integral $\int dy\,  g(y)\phi(y\varepsilon)\xrightarrow{\varepsilon\to 0}\int dy \, g(y)\phi(0) = \phi (0)$ with the test functions, produces a family of functions $1/\varepsilon \, g(x/\varepsilon)\xrightarrow{\varepsilon\to 0}  \delta(x)$. Namely, their action converges to the action of a Dirac delta $\int dx /\varepsilon \,  g(x/\varepsilon) \phi(x) = \int dy \, g(y)\phi(y\varepsilon)\xrightarrow{\varepsilon\to 0} \phi (0)$ for any test function $\phi(x)$.}. All the functions are vanishing in $x=0$,  the infinite peaks (and valleys) are just less dense around $x=0$; for any $x\neq0$ the function $\delta_\varepsilon(x)$ does not converge to anything as $\varepsilon\to 0$.

In the sudden limit $s(t)\sim\theta(t+T/2)\theta(t-T/2)$ the time integral can be computed exactly, and it reads

\begin{equation}
   \frac{2\, {\rm sin}\left( \tfrac{E_{\bf p} T}{2}\right)}{E_{\bf p}}\xrightarrow{T\to \infty} (2 \pi)\delta(E_{\bf p}).
\end{equation}
It still converges, distribution-wise, to a Dirac delta, it has a peak $\propto T$ in $0$, but the functions for the on-shell energy $E_{\bf p}>0$ are not converging to anything in the $T\to \infty$ limit, certainly not $0$. They keep bouncing periodically between the extremes $\pm 2 /E_{\bf p}$. The (fast, in the $T\to \infty$ limit) oscillations from the sin function do not grant any convergence to $0$: The Riemann-Lebesgue lemma works for functions not operators. The correct procedure is to apply the operators to a physical states, pick the partial waves amplitude (which in general are also not defined in the $T\to \infty$ limit), compute the expectation values of the observable operators (modulus squares of the partial wave functions) and see if these integrals corresponding to physical observations are well defined in the large time limit. It is interesting to note that tweaking the calculations in Section~\ref{Decay probability}, namely from time $-T/2$ to $T/2$ rather than from $0$ to $t$, one obtains the same amplitudes (hence the very same, well defined, time limits for the probabilities) as using the scattering matrix in the sudden limit. This must not be interpreted though as proof that the sudden switch is the ``correct way to insert the interaction for a finite time''. In fact, as we will show at the end of this section, there is no such a thing as a correct way to switch off the interaction asymptotically to recover the full results of the theory, not even when restricting to the long time limit behavior. The analysis of the sudden switch is to point out that if the adiabatic function $s(t)$ fulfilling~(\ref{adiabatic_0}) and~(\ref{adiabatic_1}) ``converges'' to a sudden switch (the precise meaning will be apparent later) the result of a trivial scattering matrix as in Ref.~\cite{Coleman:2018mew} can't be recovered. In order to recover the trivial scattering, one has to add to~(\ref{adiabatic_0}) and~(\ref{adiabatic_1}) the additional conditions
\begin{equation}\label{bound}
   \Delta\xrightarrow{T\to \infty} \infty,\quad \sup\left| \partial_t s(t)\right| \leq \frac{M}{\Delta}, \qquad \forall \,
 T>\bar T>0 \text{ .}
\end{equation}

In other words the modulus of the derivative of the (smooth) adiabatic function $|\dot s|$ must be smaller than a maximum that decreases at least linearly with $\Delta$ in the limit $T\to\infty$. Of course it is inessential if this bound is violated at finite times and it is required only in the large $T$ limit. One can see that manipulating directly the integral
\begin{equation}\label{Fourier}
    \int dt \, s(t) \, e^{i E_{\bf p} t} = \frac{i}{E_{\bf p}} \int dt\; \partial_t s(t) \, e^{i E_{\bf p} t} \text{ .}  
\end{equation}

The derivation by parts is granted by the smoothness of $s$ and the positivity of the on-shell energy $E_{\bf p}\geq m >0$. Thanks to the requirements~(\ref{adiabatic_0}), the time derivative of the adiabatic function is vanishing everywhere except in the intervals $t\in(-T/2-\Delta, -T/2)$ and $t\in(T/2, T/2+\Delta)$. It is convenient then to define the the functions $s_\pm$

\begin{equation}
    s_\pm (\lambda) = s\left(\pm\lambda \, \Delta  \pm \tfrac{T+\Delta}{2}\right), \qquad \lambda \in(-\tfrac{1}{2},\tfrac{1}{2})
\end{equation}
in term of the dimensionless variable $\lambda$. The integral~\eqref{Fourier} then reads

\begin{equation}
    \frac{i}{E_{\bf p}} \,{\rm exp }\left\{ -iE_{\bf p}\left(\frac{T+\Delta}{2}\right) \right\}\int d\lambda \; s^\prime_-(\lambda)\, e^{-i(E_{\bf p}\Delta)\lambda} - \frac{i}{E_{\bf p}} \,{\rm exp }\left\{ iE_{\bf p}\left(\frac{T+\Delta}{2}\right) \right\}\int d\lambda \; s^\prime_+(\lambda)\, e^{i(E_{\bf p}\Delta)\lambda},
\end{equation}

with $s^\prime_{\pm}= \partial_\lambda s_\pm$. The condition~\eqref{bound} then implies $|s^\prime_\pm|\leq M$ in the large $T$ limit. It is possible, then, to prove that both the $s_-^\prime$ and the $s^\prime_-$ integrals vanish in the large-$T$ (hence large $\Delta$) limit\footnote{For instance using the fact that any Fourier transform $\tilde{f}(\xi)=\int dx \, e^{ix\xi} f(x)$ can be written as $\tilde{f}(\xi) = \int dx \, e^{ix\xi}[f(x) - f (x +\pi/\xi)]/2$, and use the absolute convergence of the integral, the continuity and bound of $s_\pm^\prime$ at any $T>\bar{T}$ to prove that it can always be found a large enough $T$ such that the integral is arbitrary small, regardless of the specific function $s_\pm $ at said time. In other words, it vanishes in the large $T$ limit.}. The simplest and most intuitive case would be $s_-(\lambda)= s_+(-\lambda)$ independent of $T$ in the first place, them the highly oscillatory integrals vanish as a consequence of the Riemann-Lebesgue lemma. It can be seen also that, removing the bound on gradient $s_\pm^\prime$, one could have a family of functions converging to a Dirac delta\footnote{Any delta family $\delta_\varepsilon(\lambda)$ can be used as a start. It bust be multiplied by a smooth drop to the border, like $e^{-4\lambda^2/(1-4\lambda^2)}$ and properly normalized to one for all $\varepsilon$. As long as $\varepsilon\Delta\xrightarrow{T\to \infty}0$ the integrals in $s^\prime_\pm$ converge to just $1$.} $\delta(\lambda)$, substantially recovering the results of the sudden limit. Of course other outcomes are possible for adiabatic functions with unbounded derivatives in the $T\to \infty$ limit.

It is possible then to force the system to have a trivial scattering matrix\footnote{Besides a diverging phase that can be removed with a counterterm in the Lagrangian as discussed in ~\cite{Coleman:2018mew}.} even if the full interaction is on for  a long time. However it is not the case for any function (pointwise) converging $s(t)\to 1$. The smoothness of the adiabatic function can be relaxed to just continuous and smooth except a finite (constant) number of points, for instance $\{-\Delta-T/2,-T/2,T/2,\Delta+T/2\}$. One has just to be careful with the integration by parts and add ``when defined'' in every sentence containing the derivatives of $s$. On the other hand all the conditions in~\eqref{adiabatic_0},~\eqref{adiabatic_1} and~\eqref{bound} are crucial. Since the long permanence of the interaction doesn't grant a unique limit, but the behavior on the distant past and future drastically change the final outcome, one should realize that none of them in particular is supposed to reproduce the full theory. In fact none of them are.

It is possible to rework the arguments presented in this section\footnote{The calculations in Ref.~\cite{Coleman:2018mew} regarding ``Model I'' and ``Model II'' are essentially a way to resum the Dyson series. In no instance there is a switch to the momentum space, and all the calculations can be repeated without much effort integrating time in a compact domain, rather than the full real axis. In this way one obtains the $U_I(t,-T/2-\Delta)$. The formulas are the same as the one presented at the beginning of the section, except that the time integral has an upper bound in $t$, not at the end of the support of the adiabatic function $s$.} to prove the analogue of the Adiabatic theorem

\begin{equation}\label{energy_bases}
    U(t,-\tfrac{T}{2}-\Delta) |{\bf p}_1,\cdots,{\bf p}_n\rangle \sim {\rm exp}\left\{-i \left[E_0 + \sum_{j=1}^n E_{{\bf p}_j}\right] \left( t +\frac{T}{2}+\Delta \right)\right\} |{\bf p}_1,\cdots,{\bf p}_n\rangle_\Omega
\end{equation}
The states $|{\bf p}_1,\cdots,{\bf p}_n\rangle_\Omega$ build energy eigenstates basis used in the appendix~\ref{sec:app_time_dep}. The $\sim$ instead of an equality represents two facts. The first one is that it is valid in the large $T$ limit and only if all the requirements~\eqref{adiabatic_0},~\eqref{adiabatic_1} and~\eqref{bound} are fulfilled. The second is that there is an additional phase making the effective time $\Delta t_{\rm eff}\neq \Delta t= t+T/2 +\Delta$ for the $E_0$ coefficient different than the one for the $E_{{\bf p}_j}$'s; however, in the large $T$ limit $\Delta t_{\rm eff}/(t+T/2+\Delta)\to 1$. Namely, the phase is asymptotic to the correct phase.

On the other hand in the sudden case $s(t)= \theta(T/2+\Delta +t)\theta(T/2+\Delta -t)$ one has, calling $U_S$ the time evolution here to distinguish it from the previous case

\begin{equation}
\begin{split}
     &\langle{\bf q}_1,\cdots,{\bf q}_m|U_S(t,-\tfrac{T}{2}-\Delta) |{\bf p}_1,\cdots,{\bf p}_n\rangle \\
     & \quad = {\rm exp}\left\{-i \left(\sum_{j=1}^n E_{{\bf p}_j}\right) \Delta t \right\}  {}_{\Delta t}\langle{\bf q}_1,\cdots,{\bf q}_m|{\bf p}_1,\cdots,{\bf p}_n\rangle, \qquad t\in [-\tfrac{T}{2}-\Delta, \tfrac{T}{2}+\Delta]
\end{split}
\end{equation}
as it is rather obvious since the system is the exact one for $t\in (-\tfrac{T}{2}-\Delta, \tfrac{T}{2}+\Delta)$, hence time translation invariant. The results of the previous section must be recovered in this case as long as $t$ remains in its bound. needless to say, for initial conditions before $-T/2-\Delta$ or final $t>T/2+\Delta$ the phases keep growing but the $\Delta t$ in the $t$ dependent basis is stuck and the results become different from the exact ones. The difference between the two cases stems from the fact that 
\begin{equation}
    {}_{\Delta t}\langle{\bf q}_1,\cdots,{\bf q}_m|{\bf p}_1,\cdots,{\bf p}_n\rangle \neq e^{-i E_0\Delta t}\langle{\bf q}_1,\cdots,{\bf q}_m|{\bf p}_1,\cdots,{\bf p}_n\rangle_\Omega \text{ ,}
\end{equation}
something which is rather immediate to see if one works out these bases solving exactly the system, but it is rather obscure if one looks only at the Dyson series for the $U_I$: for free fields (including fields in the interaction picture) there is no difference between the basis $|{\bf p}_1,\cdots,{\bf p}_n\rangle$ at different times and the basis of the energy excitations. They are all the same basis. This is related to the fact that the regular basis of in and out states (usually assumed to exist, rather than proved, in practical applications) simply does not exist for a time independent classical source\footnote{In general, it does not exist also for time- dependent but not Fourier transformable in time sources. }. Mostly due to the fact that the state corresponding to the minimum of the Hamiltonian $\Omega\rangle$ is not a momentum eigenstate. More information about this in the Appendix~\ref{sec:app_Smatrix}. This difference between bases is crucial from the physical point of view. In the full theory, the initial and final states can be written in any basis, possibly different ones. The matrix elements of $U$ with respect to the regular $|{\bf p}_1,\cdots,{\bf p}_n\rangle$ basis are enough to write the matrix elements in all bases, one has to just use the ordinary change of basis rules. Conceptually it is fine, at worst it can be complicated numerically for inconvenient bases. If one artificially switches off the interaction in the far past and future it is not possible to distinguish, eg, if the initial or final conditions were given in the energy basis or in the particle momentum one, as they are the same. One has to ``chose a different method to insert the interaction for a long time'' to recover the exact results, if any. For instance if one has the initial and final condition in the energy basis one should use the proper adiabatic case. Because of~\eqref{energy_bases} this approach smoothly connects the free states to the energy eigenstates, then back to the corresponding free states in the far future. Seen from this perspective it is obvious that the scattering is trivial. By definition of the energy basis the only evolution of its eigenstates is a (time and energy dependent) phase. The non trivial part is that the spectrum of the Hamiltonian (besides an off-set) is the same as the free one, and with no additional degeneracy. Something that can be proved only with the ful exact solutions, as done in this paper, and cannot be obtained by looking at the Dyson series. If the initial states are given in terms of momentum (and particle number) eigenstates, the proper switch is the sudden one. If the former sends ``free states to energy eigenstates'' (which have an undefined particle number) the latter keeps the momentum and particle eigenstates (which are not energy eigenstates). It is not the purpose of this paper to consider other bases. At the moment it is also possible that for some, more complicated, bases there is not  a ``corresponding way to switch off the interaction at asymptotically large times''. In any case, there is not an universal ``correct way'' to keep the interaction for a finite, but arbitrarily large time.

\section{Conclusion}
\label{Conclusion}
An external classical source causes the particle vacuum to be unstable, and its decay into $n$ particles is given by a Poisson distribution, which is well-defined whenever the source has a Fourier transform. The time dependence of the vacuum survival probability oscillates and exhibits a quadratic behavior at short times that allows for the quantum Zeno effect. The oscillation of the survival probability dies out at large times, and it tends to a nonvanishing constant, at odds with standard particle decay. In the large volume limit the  particle density saturates, and the particle production never displays a constant rate, in contrast with the Schwinger mechanism. In addition, the precise shape of the source, and not just its effective volume, is relevant for the particle production.
\\
We find that the scattering is non trivial also for a time-independent source, the vacuum decay playing the most important role because the full dynamics can be reduced to an absorption probability amplitude of the initial particles and an interference pattern between the remaining particles and the ones produced by the vacuum.  
The result of trivial scattering was previously obtained in the framework of adiabatic switching off the interaction and under the assumption that in the limit of long persistence of the interacting Hamiltonian one can recover the full results of the unmodified theory. This claim does not hold after a closer inspection because removing the interaction changes the algebra of the physical states. In particular, the full treatment reveals the existence of a particle vacuum and an energy vacuum. As a consequence particle number eigenstates are never energy eigenstates and vice-versa.  This exact aspect of the theory cannot be recovered in the long-time limit if one switches off the interaction, no mater how it is performed. 
\\
In high energy experiments, a possible use of this specific model would be to simulate (a part of) the dilaton production out of the medium approximated as a classical source.
The effect due to the vacuum decay is fully nonperturbative and could give a relevant contribution to dilaton production, since perturbative production mechanisms are suppressed. The produced dilatons (i.e. predominantly scalar glueballs) then decay and produce mesons that can be measured, see for instance Ref. \cite{Bicudo:2012wt} where a similar process is discussed. 
\\
Any generalization of the main results of this paper to higher spin and/or charged particles would be interesting and eventually  prove useful in modeling emission particles other than the dilaton, but it remains a question whether the same techniques can be used. 

In low energy experiments, the Schwinger mechanism and Zeno effect are more easily accessible, e.g. \cite{Schmitt:2022pkd} for the Schwinger mechanism and \cite{Haroche:2014gla,2008PhRvL.101r0402B} for the Zeno effect. Therefore (appropriate extension of) the model might work as a description of those phenomena at low energies.
\\

\section*{Acknowledgments}
F. G. and A. V. acknowledge financial support from from the Polish National Science Centre (NCN) via the
OPUS project No. 2019/33/B/ST2/00613. L. T. , S.J. and A.V. have been supported by the Polish National Science Centre Grant (NCN) SONATA project No. 2020/39/D/ST2/02054.

\appendix
\section{Exact solutions of the field equations and the generators of space-time translations}
\label{sec:app_ham}
\noindent
In this appendix we show the explicit calculations for a generic classical source $g\rho(x)$, regardless of the time dependence (or independence) of $\rho$. In particular the exact solution of the field equation, and we perform the calculations for the total four-momentum operators, the Hamiltonian $\hat H$ and $\hat P$ in the Heisenberg picture.
\\
The starting Lagrangian reads

\begin{equation}\label{gen_L}
 {\cal L}= \frac{1}{2}\partial_\mu \hat \phi \; 
\partial^\mu\hat \phi -\frac{1}{2} m^2 \hat \phi^2 +g \rho \;\hat \phi,
\end{equation}
whose Euler-Lagrange equation of motion then reads
\begin{equation}\label{EL_gen}
    \Box \hat \phi + m^2\hat \phi = g\, \rho \text{ .}
\end{equation}
By hypothesis, the classical source $\rho(x)$  can be Fourier transformed with respect to the space variables, but we do not assume anything about the time dependence, beside a generic smoothness (continuous, differentiable and lack of singularities) of the real function $\rho$. In particular, we are interested in the case of a time-independent $\rho(x) =\rho({\bf x})$. An integral from $y^0=-\infty$ to a finite $t$ of $e^{\pm i p(x-y)}\rho(y)$ is not defined in general. It is surely not defined for a time-independent source. The same can be said for integrals from a finite $y^0=t$ to $+\infty$. Therefore, neither the advanced, nor the retarded green functions can be used to obtain a solution in term of a solution $\hat\phi_0$ of the homogeneous equation

\begin{equation}
    \Box \hat \phi_0 + m^2\hat \phi_0 = 0.
\end{equation}
It is possible, however, to split the time domain in two components. Without loss of generality, given a finite $t_0\in {\mathbb R}$

\begin{equation}
    (-\infty,\infty) =  (-\infty,t_0]\cup[t_0,\infty).
\end{equation}
One can safely use the advanced Green's function in the $y^0<t_0$ region, and the retarded ones in the $y^0>t_0$ region. It is immediate then to obtain the exact solutions in the the intervals $(-\infty,t_0)$ and $(t_0,\infty)$. A special solution can be written exactly in the same way (for a $t$ larger or smaller than $t_0$)
\begin{equation}
    i g \int_{t_0}^t dy^0 \int d^3y \; \rho(y)\int\frac{d^3 p}{(2\pi)^32E_{\bf p}}\left( \vphantom{\frac{}{}} e^{-i p\cdot(x-y)} - e^{i p\cdot (x-y)} \right) \text{ ,}
\end{equation}
which is manifestly continuous and differentiable in $t_0$. Therefore, the general solution reads

\begin{equation}\label{gen_sol}
    \hat \phi (x) = \hat \phi_0(x) +  i g \int_{t_0}^t dy^0 \int d^3y \; \rho(y)\int\frac{d^3 p}{(2\pi)^32E_{\bf p}}\left( \vphantom{\frac{}{}} e^{-i p\cdot(x-y)} - e^{i p\cdot (x-y)} \right).
\end{equation}
Alternatively, one can just check directly that~\eqref{gen_sol} is a solution of~\eqref{EL_gen}. Clearly the ``initial'' condition is that the free field $\hat \phi_0$ at $t=t_0$ is the full interacting field $\hat \phi(t_0,{\bf x})=\hat \phi_0(t_0,{\bf x})$. This implies that, at least for the time $t_0$, the physical field must coincide with some free field, regardless of the details of the classical source. In other words both a free field and an interacting one must have  ``the same initial conditions at $t=t_0$''.

In general, it is not granted that two fields following different equation of motions can coincide over an extended region. Namely, the Klein-Gordon equation with different sources, and the whole space at fixed time, as it is needed in Eq.~\eqref{gen_sol}.

The key to prove that~\eqref{gen_sol} is an acceptable solution is the canonical quantization rules. Indeed, they are enough to ensure that any interacting field (regardless of the equation of motion, which may be much more complicated than~\eqref{EL_gen}) coincide with some free field at fixed time in an inertial reference frame. In particular, the quantization relations ensure that, for any fixed time $\bar t$ it can be built canonically a field $\bar \phi$ which coincides with the starting field at $t=\bar t$ and fulfills the free Klein-Gordon equation. In particular at the starting time $t_0$ the field must be coinciding with some free field and $\phi_0$ can be taken as an initial condition without loss of generality.

This result, beside justifying the use of the solution~\eqref{gen_sol} in this work, is essential to use the interaction picture in the general case. More importantly, it shows that there are always some (time-dependent) ladder operators that one can use to span the particles' states for the interacting field, not only for the Lagrangian density~\eqref{gen_L}, but definitely also for the system studied in the present work. This lesser known time dependent basis is the main tool to solve exactly for the transition amplitudes in Section~\ref{Decay probability}. Therefore, as a technical reminder, we show a brief proof for the construction of such a free field and we highlight some of the consequences of this formulation.

In the case of a Hermitian scalar field $\hat\phi^\dagger=\hat \phi$, one can call ${\tilde \alpha}(x^0;p)$ and ${\tilde \alpha}^\dagger(x^0;p)$ the volume integrals

\begin{equation}\label{tilde_alpha}
    {\tilde \alpha}(x^0;p) = i\int d^3x\;  e^{ip\cdot x}\, \left[ \vphantom{\frac{}{}} \hat\Pi(x) -i E \, \hat \phi (x) \right],
\end{equation}
and its hermitian conjugate

\begin{equation}\label{tilde_alpha_dagger}
   {\tilde \alpha}^\dagger(x^0;p)= -i\int d^3x \; e^{-ip\cdot x}\left[ \vphantom{\frac{}{}} \hat\Pi(x) +i E \, \hat \phi (x) \right].
\end{equation}
The (Hermitian) field $\hat \Pi$ is the canonical conjugate of the $\hat \phi$

\begin{equation}
    \hat \Pi(t,{\bf x}) = \frac{\partial {\cal L}}{\partial(\partial_{t}\hat\phi)},
\end{equation}
which is, in turn, equal to $\hat \Pi = \partial_t \hat \phi$ in the absence of a derivative interaction. The four momentum $p=(E,{\bf p})$ at this point is generic and does not have to be on-shell. Making use of the equal-time commutators

\begin{equation}\label{eq-t_comm}
    [\hat \phi(t,{\bf x}),\hat \phi(t,{\bf y}) ]=[\hat \Pi(t,{\bf x}),\hat \Pi(t,{\bf y}) ]=0 \text{ ,} \qquad[\hat \phi(t,{\bf x}),\hat \Pi(t,{\bf y}) ] = i \delta^3({\bf x} - {\bf y}) \text{ ,}
\end{equation}
it is straightforward to check that 
\begin{equation}\begin{split}
& [\tilde{\alpha}(x^0;p_1),\tilde{\alpha}(x^0;p_2)]= [\tilde{\alpha}^\dagger(x^0;p_1),\tilde{\alpha}^\dagger(x^0;p_2)]=0,\\
&    [\tilde{\alpha}(x^0;p_1),\tilde{\alpha}^\dagger(x^0;p_2)] = e^{i(E_1-E_2)x^0} \left( \vphantom{\frac{}{}} E_1 +E_2 \right) (2\pi)^3 \delta^3({\bf p}_1 -{\bf p}_2).
    \end{split}
\end{equation}
It is immediate to note that for on-shell energies $E_1 = E_{{\bf p}_1} = \sqrt{m^2 +{\bf p}_1\cdot{\bf p}_1}$, $E_2 = E_{{\bf p}_2} = \sqrt{m^2 +{\bf p}_2\cdot{\bf p}_2}$ this is a generalization of the commutation relations of the ladder operators. More precisely, for an uncharged free field $\hat \phi_f$ with mass $m$ and normalization

\begin{equation}
    \hat \phi_f (x) = \int\frac{d^3 p}{(2\pi)^3\sqrt{2 E_{\bf p}}} \left[ \vphantom{\frac{}{}} a_f({\bf p}) e^{-ip \cdot x} + a^\dagger_f({\bf p}) e^{ip\cdot x} \right],
\end{equation}
one has
\begin{equation}\label{free_comm}
    [a_f({\bf p}), a_f({\bf q})]=[a^\dagger_f({\bf p}), a^\dagger_f({\bf q})]=0,\qquad [a_f({\bf p}), a^\dagger_f({\bf q})] = (2\pi)^3 \delta^3 ({\bf p} -{\bf q}) \text { ,}
\end{equation}
therefore, the on-shell $\alpha(x^0;m,{\bf{p}})= \tilde\alpha(x^0; E_{\bf p},{\bf p})$ is a time dependent generalization of the free $\sqrt{2E_{\bf p}}a_f({\bf p})$ to the interacting case. In other words 
\begin{equation}\label{a_p_t}
    a_{\bf p}(x^0,m)= \frac{\alpha(x^0;m,{\bf p})}{\sqrt{2E_{\bf p}}} = \frac{\tilde\alpha(x^0; E_{\bf p},{\bf p})}{\sqrt{2E_{\bf p}}}, \qquad  a^\dagger_{\bf p}(x^0,m)= \frac{\alpha^\dagger(x^0;m,{\bf{p}})}{\sqrt{2E_{\bf p}}} = \frac{\tilde\alpha^\dagger(x^0; E_{\bf p},{\bf p})}{\sqrt{2E_{\bf p}}},
\end{equation}
are the generalization, respectively, of the destruction and creation operators. They have the exactly the same commutation relations as the free ones~\eqref{free_comm}. Therefore, if one has an uncharged interacting field $\phi$ and its canonical conjugate $\Pi$ at the same time, it is possible to write both of them in terms of the $a_{\bf p}(x^0,m)$ and $a_{\bf p}^\dagger(x^0,m)$ operators
\begin{equation}\label{cfr_phi-Pi}
    \begin{split}
            \hat \phi (x^0,{\bf x}) &= \int\frac{d^3 p}{(2\pi)^3\sqrt{2 E_{\bf p}}} \left[ \vphantom{\frac{1}{2}} a_{\bf p}(x^0,m) \; e^{-iE_{\bf p} x^0} e^{ i{\bf p}\cdot {\bf x}} + a^\dagger_{\bf p}(x^0,m) \; e^{iE_{\bf p} x^0} e^{ -i{\bf p}\cdot{\bf x}} \right],\\
            \hat \Pi (x^0,{\bf x}) &= -i\int\frac{d^3 p}{(2\pi)^3\sqrt{2 E_{\bf p}}} \, E_{\bf p}\left[ \vphantom{\frac{1}{2}} a_{\bf p}(x^0,m) \; e^{-iE_{\bf p} x^0} e^{ i{\bf p}\cdot {\bf x}} - a^\dagger_{\bf p}(x^0,m) \; e^{iE_{\bf p} x^0} e^{ -i{\bf p}\cdot{\bf x}} \right],
    \end{split}
\end{equation}
which can be directly verified by simply plugging the definitions~\eqref{a_p_t} and then~\eqref{tilde_alpha} and ~\eqref{tilde_alpha_dagger} in the integrals. The last equations clearly work exclusively for the same time $x^0$ in the exponential and in the time-dependent ladder operators.

Finally, if one builds the (normal ordered\footnote{The normal ordering is defined in the same way as for the free fields. This definition holds in the interacting case thanks to~\eqref{exact_field}.}) Klein-Gordon Hamiltonian with the fields at time $x_0$ and the same mass as the one in the definition of $a_{\bf p}(x^0, m)$
\begin{equation}\label{:H0:}
    \hat H_0 = \frac{1}{2}:\!\!\!\int d^3x \left[ \hat\Pi^2(x^0,{\bf x}) + \nabla_{\bf x}\hat \phi(x^0,{\bf x})\cdot  \nabla_{\bf x}\hat\phi(x^0,{\bf x}) + m^2  \hat\phi(x^0,{\bf x})\hat\phi(x^0,{\bf x})\right]\!\!: \text { ,}
\end{equation}
it is possible to prove, making use of the commutation relations~\eqref{eq-t_comm} and the definitions~\eqref{cfr_phi-Pi},~\eqref{a_p_t},~\eqref{tilde_alpha} and \eqref{tilde_alpha_dagger}, that the commutators of the generalized ladder operators with $\hat H_0$ are the same as the commutators of a free field
\begin{equation}\label{comm_H0}
    [\,\hat H_0\;, \; a_{\bf p}(x^0,m)\,] = -E_{\bf p}\;a_{\bf p}(x^0,m) \text{ ,} \qquad  [\,\hat H_0 \;, \; a^\dagger_{\bf p}(x^0,m) \,] = E_{\bf p}\; a^\dagger_{\bf p}(x^0,m) \text { .}
\end{equation}
Therefore, using the free propagator ${\rm exp}\{-i\hat H_0\Delta x^0\}$, it is possible to write a free field $\bar \phi$ that follows the free Klein-Gordon equation with mass $m$ and coincides with the physical field at a specific time $\bar t$, $\bar \phi_0({\bar t},{\bf x})=\bar \phi({\bar t},{\bf x})$ : 
\begin{equation}\label{phi_0}
\begin{split}
    {\bar \phi} (t,{\bf x}) &= e^{i\hat H_0(t-{\bar t})} \; \hat \phi({\bar t},{\bf x}) \; e^{-i\hat H_0(t-{\bar t})} \\
    &= \int\frac{d^3 p}{(2\pi)^3\sqrt{2 E_{\bf p}}} \left[ \vphantom{\frac{1}{2}} a_{\bf p}({\bar t},m) \; e^{-iE_{\bf p} t} e^{ i{\bf p}\cdot {\bf x}} + a^\dagger_{\bf p}({\bar t},m) \; e^{iE_{\bf p} t} e^{ -i{\bf p}\cdot{\bf x}} \right],
\end{split}
\end{equation}
as it was stated before. The same result holds for the analogous $\bar  \Pi$ that coincides with $\hat\Pi$ at  ${\bar t}$ and evolves according to the free Klein-Gordon equation.

These arguments are rather general, so they work, in particular, for the simple Lagrangian density in~\eqref{gen_L}. Without loss of generality, for any $t_0$ one can build a free field that coincides with the full interacting field at $t_0$. The solution in~\eqref{gen_sol} is thus general. 

It is possible now to plug~\eqref{phi_0}, with $t=t_0$, in~\eqref{gen_sol}. Writing all in therm of the same $d^3p$ integral and factoring the ${\rm exp}\{\pm i p\cdot x\}$ phases from both of the terms

\begin{equation}\label{gen_sol_simp}
    \hat\phi(x) = \int \frac{d^3 p}{(2\pi)^3\sqrt{2 E_{\bf p}}} \left[ \vphantom{\frac{1}{2}} \left(  \vphantom{\frac{}{}} a_{\bf p}(t_0, m) + f(t,t_0;E_{\bf p},{\bf p})  \right) e^{-ip\cdot x} + \left(  \vphantom{\frac{}{}} a^\dagger_{\bf p}(t_0, m) + f^*(t,t_0;E_{\bf p},{\bf p})  \right) e^{ip\cdot x} \right],
\end{equation}
the function $f$ being the integral of the classical source

\begin{equation}\label{gen_f}
    f(t,t_0;p) = i\, g \int_{t_0}^t d y^0\int d^3 y \; \frac{\rho(y)}{\sqrt{2 E_{\bf p}}} \; e^{i p\cdot y},
\end{equation}
and $f^*$ its complex conjugate. In almost all the cases in the manuscript, including~\eqref{gen_sol_simp}, only the on-shell integral is needed. For the most part of the paper, also, only a time-independent $\rho$ is considered and the $y^0$ integral can be performed exactly. For ease of reading in the main text the the on-shellness is understood and the $t_0$  dependence (taken as $t_0=0$) is omitted. In a similar manner the formal notation $a_{\bf p}(t_0=0,m)$ is left out in favor of the simpler $a_{\bf p}$. The exact field $\hat\Pi= \partial_t \hat \phi$ can be computed directly from from~\eqref{gen_sol_simp}. 

\begin{equation}\label{Pi_gen}
\begin{split}
    \hat \Pi (x) = \partial_t\hat\phi(x) = -i\int \frac{d^3 p}{(2\pi)^3\sqrt{2 E_{\bf p}}} E_{\bf p}&\left[ \left(  \vphantom{\frac{}{}} a_{\bf p}(t_0, m) + f(t,t_0;E_{\bf p},{\bf p}) +\frac{i}{E_{\bf p}} \partial_tf(t,t_0;E_{\bf p},{\bf p})  \right) e^{-ip\cdot x} \right. \\
    &\left. - \left(  \vphantom{\frac{}{}} a^\dagger_{\bf p}(t_0, m) + f^*(t,t_0;E_{\bf p},{\bf p})  -\frac{i}{E_{\bf p}} \partial_tf^*(t,t_0;E_{\bf p},{\bf p}) \right) e^{ip\cdot x} \right].
\end{split}
\end{equation}
Making use of the definition~\eqref{gen_f} it is immediate to show that, for any on-shell and positive frequency four-momentum $p^\mu$

\begin{equation}\label{gen-rel_fpoint}
   \begin{split}
       \partial_t f^*(t,t_0;E_{\bf p},{\bf p}) \; e^{i(E_{\bf p})t}\,e^{-i{\bf p}\cdot{\bf x}} &= -i\, g \int d^3 y \; \frac{\rho(t,{\bf y})}{\sqrt{2 E_{\bf p}}} \; e^{-i {\bf p}\cdot ({\bf x} -{\bf y})}= -\left( i\, g \int d^3 y \; \frac{\rho(t,{\bf y})}{\sqrt{2 E_{\bf p}}} \; e^{i (-{\bf p})\cdot ({\bf x} -{\bf y})}  \right) \\
       &= - \partial_t f(t,t_0;E_{\bf -p}, {\bf -p}) \; e^{-i(E_{-{\bf p}})t } \, e^{ i(-{\bf p})\cdot{x}}.
   \end{split} 
\end{equation}
Within the minimal integrability hypothesis for the classical source (and its partial Fourier transform), the two terms in the time derivatives of $f$ in~\eqref{Pi_gen} converge separately, and they have an opposite sign, as it is apparent performing a ${\bf p} \to -{\bf p}$ in one of them and not the other. Their contribution to the total field simplifies

\begin{equation}\label{Pi_gen_simp}
\begin{split}
    \hat \Pi (x) =  -i\int \frac{d^3 p}{(2\pi)^3\sqrt{2 E_{\bf p}}} E_{\bf p}&\left[ \left(  \vphantom{\frac{}{}} a_{\bf p}(t_0, m) + f(t,t_0;E_{\bf p},{\bf p})   \right) e^{-ip\cdot x} - \left(  \vphantom{\frac{}{}} a^\dagger_{\bf p}(t_0, m) + f^*(t,t_0;E_{\bf p},{\bf p}) \right) e^{ip\cdot x} \right].
\end{split}
\end{equation}
Confronting the results~\eqref{gen_sol_simp} and~\eqref{Pi_gen_simp} with the general formulas~\eqref{cfr_phi-Pi}, it is immediate to recognize

\begin{equation}\label{a(t)_exact}
    a_{\bf p}(t_0,m) + f(t,t_0;E_{\bf p},{\bf p}) = a_{\bf p}(t,m), \qquad a^\dagger_{\bf p}(t_0,m) + f^*(t,t_0;E_{\bf p},{\bf p}) = a^\dagger_{\bf p}(t,m),
\end{equation}
as the on-shell ladder operators for the generic time $t$. Hence, the number operator in the Heisenberg picture reads, in term of the ladder operators at $t_0$, and the integrals of the classical source

\begin{equation}\label{N_gen}
    \hat N(t) = \int \frac{d^3 p}{(2\pi)^3}\;  a^\dagger_{\bf p}(t, m) a_{\bf p}(t, m) = \int \frac{d^3 p}{(2\pi)^3} \left( \vphantom{\frac{}{}} a^\dagger_{\bf p}(t_0,m) + f^*(t,t_0;E_{\bf p},{\bf p})\right) \left( \vphantom{\frac{}{}} a_{\bf p}(t_0,m) + f(t,t_0;E_{\bf p},{\bf p})\right).
\end{equation}
It is possible then to compute now the total four-momentum operators. The system is, in general,  not space-time translation invariant, but $\hat H(t)$ and $\hat {\bf P}(t)$ are still the generators of space-time translations in the lab frame at time $t$. They are the regular volume integral of the (contribution of the quantum field to the total) stress-energy tensor $\hat T^{\mu\nu}$, for a non-derivative interaction term like~\eqref{gen_L} this has the standard form

\begin{equation}
    \hat T^{\mu\nu}(x) =\frac{1}{2}\partial^\mu \hat \phi(x) \partial^\nu \hat \phi(x) + \frac{1}{2}\partial^\nu \hat \phi(x) \partial^\mu \hat \phi(x) - g^{\mu\nu}{\cal L},
\end{equation}
 and thus
\begin{equation}\label{gen_4_mom_op}
    \begin{split}
        \hat H(t) & = \int d^3 x \; \hat T^{00} =  \frac{1}{2}\int d^3 x \left[ \hat \Pi (x)^2 + \nabla_{\bf x} \hat \phi\cdot \nabla_{\bf x} \hat \phi (x) + m^2\hat \phi (x)^2 - 2\, g\, \rho(x)\, \hat \phi(x)\right] \\
        \hat P^i(t) & = \int d^3 x \; \hat T^{0i} =  \frac{1}{2}\int d^3 x \left[ \hat \Pi(x) \; \partial^i \hat \phi(x) \,+\, \partial^i \hat \phi(x) \; \hat \Pi(x) \right] .
    \end{split}
\end{equation}
The derivation is usually mentioned in textbook~\cite{Gelis:2019yfm,Coleman:2018mew}. Also, the computations of the volume integrals follows the steps of the textbook examples for the free scalar field since the~\eqref{gen_4_mom_op} involves only $\hat \phi$, $\hat \Pi$ and the spatial gradients $\nabla_{\bf x}\hat\phi$ at equal limes, and they all can be written using the same algebra of the free fields~\eqref{cfr_phi-Pi}, including the normal ordering. The notable difference is the  additional term due to the interaction with the classical source. The time-dependent momentum operator in the Heisenberg picture then reads

\begin{equation}\label{gen_P}
    :\!\!\hat{\bf P}(t)\!\!: = \int\frac{d^3 p}{(2\pi)^3} \; {\bf p} \; \left( \vphantom{\frac{}{}} a^\dagger_{\bf p}(t_0,m) + f^*(t,t_0;E_{\bf p},{\bf p})\right) \left( \vphantom{\frac{}{}} a_{\bf p}(t_0,m) + f(t,t_0;E_{\bf p},{\bf p})\right),
\end{equation}
it is immediate to verify then, since the commutators of the time-dependent ladder operator are the same as the free ones, that this is a valid generator of the space-translations

\begin{equation}
    [:\!\!\hat{ P}^j(t)\!\!:, \hat \phi (t,{\bf x})] = -\int\frac{d^3 p}{(2\pi)^3\sqrt{2E_{\bf p}}} \; p^j \,\left[ \vphantom{\frac{}{}} a_{\bf p}(t, m) \, e^{-ip\cdot x} - a^\dagger_{\bf p}(t,m)\, e^{ip\cdot x}  \right] = i \partial^j \hat \phi (t,{\bf x}) = -i\partial_j \hat \phi(t, {\bf x}).
\end{equation}
In other words, the fields at time $t$ translate in space with the unitary operator ${\rm exp}\{ i :\hat{\bf P}(t):\; \cdot \; \Delta{\bf x} \}$. Since the interaction in the Lagrangian density~\eqref{gen_L} is so simple, the additional term in the Hamiltonian can be easily computed in terms of the on-shell $f$ function~\eqref{gen_f}. The normal ordering is also trivial

\begin{equation}\begin{split}
    :\!\!\hat H(t)\!\!: &= \int\frac{d^3 p}{(2\pi)^3} \; E_{\bf p} \; a^\dagger_{\bf p}(t,m) \, a_{\bf p}(t,m) \\
    & \qquad +\int\frac{d^3 p}{(2\pi)^3} \left[ \vphantom{\frac{}{}}-i\partial_t f^*(t,t_0;E_{\bf p}, {\bf p})\;  a_{\bf p}(t, m) +i\partial_t f(t,t_0;E_{\bf p}, {\bf p})\;  a^\dagger_{\bf p}(t, m)  \right],
\end{split}\end{equation}
which is convenient to reorganize in a free-like operatorial part, and a time-dependent offset (function, not an operator)

\begin{equation}\label{gen_H}
    \begin{split}
        :\!\!\hat H(t)\!\!: &= \int\frac{d^3 p}{(2\pi)^3} \; E_{\bf p}\, \left( a^\dagger_{\bf p}(t,m) -\frac{i}{E_{\bf p}} \partial_t f^*(t,t_0;E_{\bf p},{\bf p}) \right)\left( a_{\bf p}(t,m)  +\frac{i}{E_{\bf p}} \partial_t f(t,t_0;E_{\bf p},{\bf p}) \right) \\
        & \qquad -\int\frac{d^3 p}{(2\pi)^3}\; E_{\bf p} \,\left| \frac{\partial_t f(t,t_0;E_{\bf p},{\bf p})}{E_{\bf p}} \right|^2.
    \end{split}
\end{equation}
The commutation relations of the $a_{\bf p}(t,m)$ and $a^\dagger_{\bf p}(t,m)$ are, by construction, the same as the free ones~\eqref{free_comm}, however the shift in the operators by a function in~\eqref{gen_H} produces a change in the commutation relations with the ladder operators. The Hamiltonian is no longer an integral of $Ea^\dagger a$ at the operator level,  but of $Ea^\dagger a -i f^* a + ifa^\dagger$ plus functions (that commute with everything). Then

\begin{equation}\label{com_time-dep_H}
    \begin{split}
        [:\!\!\hat H(t)\!\!:, a_{\bf p}(t, m)] &= -E_{\bf p} a_{\bf p} (t,m) - i\partial_t f(t,t_0;E_{\bf p},{\bf p}), \\
        [:\!\!\hat H(t)\!\!:, a^\dagger_{\bf p}(t, m)] &= E_{\bf p} a^\dagger_{\bf p} (t,m) - i\partial_t f^*(t,t_0;E_{\bf p},{\bf p}).
    \end{split}
\end{equation}
If one takes the commutator of $i[:\!\! H (t)\!\!:,\hat \phi(t,{\bf x})]$, the extra functions reconstruct the the field $\hat \Pi(t,{\bf x})$ as in~\eqref{Pi_gen}, which can be simplified thanks to~\eqref{gen-rel_fpoint} in~\eqref{Pi_gen_simp} as shown before. The extra functions do not simplify if one takes the commutator with $\hat \Pi (t,{\bf x})$, but they are needed to obtain the correct time derivative $\partial_t\hat \Pi(t,{\bf x})$. The higher order derivatives need the second (and higher) time derivatives of $f$. They are given not only by the nested commutators with $:\!\! H (t)\!\!:$, but they involve also commutators with the time derivatives of $:\!\! H (t)\!\!:$. Rather than making some specific examples it is easier to sum all up and and remind that the fields evolve in time with the unitary propagator $U(t^\prime,t)$ that solves the time-dependent Shr\"odinger equation w.r.t. the variable $t^\prime$:
\begin{equation}\label{gen_prop}
    \begin{split}
        i\partial_{t^\prime} \; U(t^\prime,t) & =\; :\!\!\hat H_S(t^\prime)\!\!: \; U(t^\prime,t) =  U(t^\prime,t)U^{-1}(t^\prime,t) \; :\!\!\hat H_S(t^\prime)\!\!: \; U(t^\prime,t) \\
        &= U(t^\prime,t){ U}^{\dagger}(t^\prime,t) \; :\!\!\hat H_S(t^\prime)\!\!: \; U(t^\prime,t) = { U}(t^\prime,t) \; :\!\!\hat H(t^\prime)\!\!: \; \text{ .}
    \end{split}
\end{equation}
Some comments about Eq. ~\eqref{gen_prop} are in order. Since it is standard textbook material, we don't think it is necessary to give too many explanations. It is important to note though, since it is a common source of confusion, that the Hamiltonian is on the left if one uses the Shr\"odinger time-dependent operator and, by construction, on the right if one uses (as it is common in quantum field theories) the Hamiltonian in the Heisenberg picture. The propagator is the same in both pictures, however their Dyson series needs the regular time ordering if the equation is solved for the $\hat H_S(t)$ and anti-time-ordering if solved for the Heisenberg one $\hat H(t)$. In the case of time independent operators the time ordering is irrelevant $\hat H_S=\hat H$, the propagator is a regular exponential and it commutes with the Hamiltonian, it doesn't matter if it is on the right or on the left. It bears mentioning that in the interaction picture one starts with the operators in the Shr\"odinger picture (including the Hamiltonian) and then evolves them with the free Hamiltonian to get their interaction picture counterpart.  

Since the field can be written in the form

\begin{equation}
    \hat \phi(t^\prime,{\bf x}) = U^\dagger(t^\prime, t) \; \hat \phi(t,{\bf x}) \; U(t^\prime, t),
\end{equation}
the $n-$th time derivative in $t$ read
\begin{equation}
    \left( \partial_t \right)^n \hat \phi(t,{\bf x}) = \lim_{t^\prime\to t} \left[  \left( \partial_{t^\prime} \right)^n \hat \phi(t^\prime,{\bf x}) \right] = \lim_{t^\prime\to t}\left\{\left(\partial_{t^\prime}\right)^n \left[ U^\dagger(t^\prime, t) \; \hat \phi(t,{\bf x}) \; U(t^\prime, t) \right]\right\}.
\end{equation}
The dependence in $t^\prime$ in the right hand side is only on the propagators. Therefore
\begin{equation}
    \partial_t \hat \phi(t,{\bf x}) = i[:\!\! \hat H(t)\!\!:\, , \, \hat \phi(t,{\bf x})], 
\end{equation}
the second time derivative reads
\begin{equation}
    \partial^2_t  \hat \phi(t,{\bf x}) = i[\partial_t:\!\! \hat H(t)\!\!:\, , \, \hat \phi(t,{\bf x})] -\left[\,:\!\! \hat H(t)\!\!:\, , [:\!\! \hat H(t)\!\!:\, , \, \hat \phi(t,{\bf x})]\, \right] \text{ .}
\end{equation}
As it can be shown using~\eqref{gen_f}, the first commutator vanishes by construction.  But for the higher order derivatives, nested commutators involving the Hamiltonian and its partial derivatives, the formula cannot be reduced to the one valid in the time-independent case that involves only $:\!\!\hat H\!\!:$. It is possible also to verify order by order that these formulas are equivalent to just taking the time derivatives of the explicit solution~\eqref{gen_sol_simp}. 
\\
The last worthwhile calculation here is the commutators between the time-dependent Hamiltonian and momentum operators. Making direct use of~\eqref{com_time-dep_H}
\begin{equation}\label{Poin_viol}
    \begin{split}
       -i\partial_t:\!\! \hat {\bf P}(t)\!\!: \; = [:\!\! \hat H(t)\!\!: \, , \, :\!\! \hat {\bf P}(t)\!\!:] \, = -i\int\frac{d^3 p}{(2\pi)^3} \, {\bf p}\left[ \vphantom{\frac{}{}} \partial_t f^*(t,t_0; E_{\bf p},{\bf p}) \,  a_{\bf p}(t,m) \; + \; \partial_t  f(t,t_0;E_{\bf p},{\bf p}) \,  a^\dagger_{\bf p}(t,m) \right] \text{ .}
    \end{split}
\end{equation}

This is in general different from $0$. For a generic $\rho$ it might be possible to have commuting operators at some times. However in the time-independent case $\rho = \rho({\bf x})$, it can never happen. The function
\begin{equation}
   f(t,t_0;E_{\bf p},{\bf p}) +\frac{i}{E_{\bf p}}\partial_t f(t,t_0;E_{\bf p},{\bf p}) = -\frac{g \; e^{i E_{\bf p}t_0}}{\sqrt{2E_{\bf p}^3}} \int d^3 x \; \rho({\bf x}) e^{-i{\bf p}\cdot x},
\end{equation}
as well as the modulus square of
\begin{equation}
   \partial_t f(t,t_0;E_{\bf p},{\bf p}) = \frac{ig \; e^{i E_{\bf p}t}}{\sqrt{2E_{\bf p}}} \int d^3 x \; \rho({\bf x}) e^{-i{\bf p}\cdot x},
\end{equation}
are time-independent, making the Hamiltonian time-independent too. As well as making the operator in the right-hand side of~\eqref{Poin_viol} never trivial.
In the rest of the manuscript, the time ordering is understood, and the $:\cdots:$ is omitted.

\section{Phase of the vacuum overlap}
\label{sec:app_time_dep}
In \eqref{mod} we have found the modulus of the vacuum overlap. It remains to verify that the phase of the overlap ${}_t\langle 0|0\rangle$ is given by \eqref{phase phi}. In order to do so, it is useful to introduce a basis of energy eigenstates. The procedure is substantially the same as the one employed in Section~\ref{Decay probability}, for the time-dependent basis~\eqref{t_states}. Looking at the generic, time-dependent Hamiltonian~\eqref{gen_H}; one recognizes the ``energy ladder operators'' $b_{\bf p}(t)$ and $b^\dagger_{\bf p}(t)$

\begin{equation}
    b_{\bf p}(t) = a_{\bf p}(t) +\frac{i}{E_{\bf p}} \dot f(t,{\bf p}) = a_{\bf p} + f(t,{\bf p})+\frac{i}{E_{\bf p}} \dot f(t,{\bf p}),
\end{equation}
and their Hermitian conjugate for the creation operators. In the latter formula it has been used the compact notation

\begin{equation}
    \dot f(t,{\bf p}) = \partial_t f(t,0;E_{\bf p},{\bf p}).
\end{equation}
In particular, for a time-independent source $b_{\bf p} = a_{\bf p} + h({\bf p})$ which is time independent. For the rest of this Appendix we will consider the general case. The time independent one being just a particular subset.

The complete basis of energy eigenstates can be obtained as it has been done in the previous case. The energy vacuum reads

\begin{equation}\label{Omega_t}
    \begin{split}
        |\Omega(t)\rangle &= {\rm exp}\left\{-\frac{1}{2} \int \frac{d^3 p}{(2\pi)^3} \left| f(t,{\bf p})+\frac{i}{E_{\bf p}} \dot f(t,{\bf p})\right|^2\right\} {\rm exp}\left\{- \int \frac{d^3 p}{(2\pi)^3} \left( f(t,{\bf p})+\frac{i}{E_{\bf p}} \dot f(t,{\bf p})\right) b^\dagger_{\bf p}(t)\right\} |0\rangle\\
        &\Rightarrow b_{\bf p}(t) \, |\Omega(t)\rangle = 0, \quad \hat H(t)|\Omega(t)\rangle = E_0(t)|\Omega(t)\rangle, \qquad E_0(t) =-\int\frac{d^3 p}{(2\pi)^3 } \, E_{\bf p} \,\left| \frac{\dot f(t,{\bf p})}{E_{\bf p}} \right|^2<0,
    \end{split}
\end{equation}
and therefore

\begin{equation}\label{more_coh}
    \begin{split}
       & a_{\bf p}(t) \, |\Omega(t)\rangle = -\frac{i}{E_{\bf p}} \dot f(t,{\bf p})  \, |\Omega(t)\rangle, \qquad a_{\bf p} \, |\Omega(t)\rangle = -\left(  f(t,{\bf p})+\frac{i}{E_{\bf p}} \dot f(t,{\bf p})\right)|\Omega(t)\rangle \\
       & b_{\bf p}(t) \, |0\rangle_t = \frac{i}{E_{\bf p}} \dot f(t,{\bf p})  \, |0\rangle_t, \qquad b_{\bf p}(t) \, |0\rangle = \left(  f(t,{\bf p})+\frac{i}{E_{\bf p}} \dot f(t,{\bf p})\right)|0\rangle.
    \end{split}
\end{equation}
The excited states are build in the usual way

\begin{equation}\label{exc_Omega_t}
    \begin{split}
        |{\bf p}_1\,\cdots\, {\bf p}_n \rangle_{\Omega(t)} &= \sqrt{\frac{2E_{{\bf p}_1}\cdots 2E_{{\bf p}_n}}{n!}} \;  b^\dagger_{{\bf p}_1}(t) \cdots b^\dagger_{{\bf p}_n}(t) \, |\Omega(t)\rangle, \\
        & \Rightarrow \hat H(t) \,  |{\bf p}_1\,\cdots\, {\bf p}_n \rangle_{\Omega(t)} = \left( E_0(t) +\sum_{j=1}^n E_{{\bf p}_j} \right)  |{\bf p}_1\,\cdots\, {\bf p}_n \rangle_{\Omega(t)} \; .
    \end{split}
\end{equation}
Inserting the identity in terms of the energy eigenstates at time $t$, it is possible to find an identity that will be useful later

\begin{equation}\label{formula}
    \begin{split}
        {}_t\langle 0|0\rangle &= {}_t\langle 0| \, {\mathds 1} \, |0\rangle = \sum_{n=0}^\infty \left[ \prod_{j=1}^n \int\frac{d^3 p_j}{(2\pi)^3 2 E_{{\bf p}_j}} \right] {}_t\langle 0|{\bf p}_1\,\cdots\, {\bf p}_n  \rangle_{\Omega(t)}\, {}_{\Omega(t)}\langle {\bf p}_1\,\cdots\, {\bf p}_n|0\rangle \\
        &= \sum_{n=0}^\infty\frac{1}{n!}\left[ \prod_{j=1}^n \int\frac{d^3 p_j}{(2\pi)^3 } \right] {}_t\langle 0|\, b^\dagger_{{\bf p}_1}(t) \cdots b^\dagger_{{\bf p}_n}(t) \, |\Omega(t)\rangle\langle\Omega(t)| \, b_{{\bf p}_1}(t) \cdots b_{{\bf p}_n}(t) \,|0\rangle \\
        &=  {}_t\langle 0|\Omega(t)\rangle\,\langle\Omega(t)|0\rangle \sum_{n=0}^\infty\frac{1}{n!}\left[ \int\frac{d^3 p}{(2\pi)^3} \left( -\frac{i}{E_{\bf p}}\dot f^*(t,{\bf p}) \right)\left( f(t, {\bf p})  +\frac{i}{E_{\bf p}}\dot f(t,{\bf p})\right) \right]^n \\
        &= {}_t\langle 0|\Omega(t)\rangle\,\langle\Omega(t)|0\rangle\; {\rm exp} \left\{ \int\frac{d^3 p}{(2\pi)^3} \left( -\frac{i}{E_{\bf p}}\dot f^*(t,{\bf p}) \right)\left( f(t, {\bf p})  +\frac{i}{E_{\bf p}}\dot f(t,{\bf p})\right) \right\}.
    \end{split}
\end{equation}
On the other hand, making use of~\eqref{mod}, without loss of generality

\begin{equation}\label{varphi}
    {}_t\langle 0|0\rangle = e^{-\frac{1}{2}N_0(t) +i\varphi(t)}, \qquad \Rightarrow e^{i\varphi(t)} = e^{\frac{1}{2}N_0(t)}{}_t\langle 0|0\rangle,
\end{equation}
and making use of this, the definition~\eqref{t_states} and the Shr\"odinger equation in the Heisenberg representation~\eqref{gen_prop}

\begin{equation}\label{varphidot}
    \begin{split}
        {}_t\langle 0|0\rangle &= \langle 0|\, U_I(t,0) \,|0\rangle = \langle 0|\, U(t,0) \,|0\rangle\\
        \Rightarrow & i \, \dot \varphi(t) \; e^{i\varphi(t)} = \frac{1}{2}\dot N_0(t)\; e^{\frac{1}{2}N_0(t)}{}_t\langle 0|0\rangle -i e^{\frac{1}{2}N_0(t)}\langle 0|\,U(t,0) \; \hat H(t)\,|0\rangle.
    \end{split}
\end{equation}
Calling ${\cal I}(t)$ the integral

\begin{equation}\label{I(t)}
    {\cal I}(t) = \int\frac{d^3 p}{(2\pi)^3} \left( -\frac{i}{E_{\bf p}}\dot f^*(t,{\bf p}) \right)\left( f(t, {\bf p})  +\frac{i}{E_{\bf p}}\dot f(t,{\bf p})\right),
\end{equation}
and making use of the identity~\eqref{formula} and of~\eqref{Omega_t},~\eqref{exc_Omega_t},~\eqref{more_coh} one can write

\begin{equation}
    \begin{split}
        &\langle 0|\,U(t,0) \; \hat H(t)\,|0\rangle = {}_t\langle 0| \, \hat H(t)\,|0\rangle = \sum_{n=0}^\infty\left[ \prod_{j=1}^n \int\frac{d^3 p_j}{(2\pi)^32E_{{\bf p}_j}} \right]{}_t\langle 0|{\bf p}_1, \cdots {\bf p}_n\rangle_{\Omega(t)} \; {}_{\Omega(t)}\langle {\bf p}_1, \cdots {\bf p}_n |\, \hat H(t) \,|0\rangle \\
        &= {}_t\langle 0|\Omega(t)\rangle\,\langle\Omega(t)|0\rangle\sum_{n=0}^\infty\frac{1}{n!}\left[ \prod_{j=1}^n \int\frac{d^3 p_j}{(2\pi)^3} \left( -\frac{i}{E_{{\bf p}_j}}\dot f^*(t,{\bf p}_j) \right)\left( f(t, {\bf p}_j)  +\frac{i}{E_{{\bf p}_j}}\dot f(t,{\bf p}_j)\right) \right] \left( E_0(t) +\sum_{j=1}^nE_{{\bf p}_j} \right)\\
        &={}_t\langle 0|\Omega(t)\rangle\,\langle\Omega(t)|0\rangle \, {\rm exp}\left\{{\cal I}(t)\vphantom{\frac{}{}}\right\} \, \left[ E_0(t) +  \int\frac{d^3 p}{(2\pi)^3} \left( -\frac{i}{E_{{\bf p}}}\dot f^*(t,{\bf p}) \right)\left( f(t, {\bf p})  +\frac{i}{E_{{\bf p}}}\dot f(t,{\bf p})\right) E_{\bf p}\right]\\
        &=  {}_t\langle 0|0\rangle \,  \int\frac{d^3 p}{(2\pi)^3} \, \left[-i \vphantom{\frac{}{}} \dot f^*(t,{\bf p})\, f(t, {\bf p}) \right].
    \end{split}
\end{equation}
Plugging into Eq.\eqref{varphidot}, and making use of~\eqref{varphi} to simplify the $e^{i\varphi}$ phase

\begin{equation}
    \dot \varphi(t) = -\frac{i}{2} \dot N_0(t) + i \int\frac{d^3 p}{(2\pi)^3} \, \dot f^*(t,{\bf p})\, f(t, {\bf p}) = \frac{i}{2} \int\frac{d^3 p}{(2\pi)^3} \left[ \dot f^*(t,{\bf p}) \, f(t,{\bf p}) - f^*(t,{\bf p}) \, \dot f(t,{\bf p}) \vphantom{\frac{}{}} \right].
\end{equation}
Since the phase of $\langle0|0\rangle$ is $0$ by construction, the latter is enough to prove that Eq.~\eqref{phase phi} is the correct solution, for time-dependent classical source in general, and in particular the time-independent source too. The last step is to verify that is a real number as expected. This can be done making direct use of the definition of the $f$ function~\eqref{gen_f}

\begin{equation}
    \begin{split}
        \varphi(t) &= \frac{i}{2} \int_0^t dx^0\int\frac{d^3 p}{(2\pi)^3} \left[ \partial_{x^0} f^*(x^0,{\bf p}) \, f(x^0,{\bf p}) - f^*(x^0,{\bf p}) \, \partial_{x^0} f(x^0,{\bf p}) \vphantom{\frac{}{}} \right] \\
        &=\frac{i\, g^2}{2}\int\frac{d^3 p}{(2\pi)^32E_{\bf p}}\int_0^tdx^0\int_0^{x_0}dy^0\int d^3 x\int d^3y \,  \rho(x) \, \rho(y)\left[ e^{-ip\cdot(x-y)} -e^{ip\cdot(x-y)}\right]\\
        &= g^2 \int\frac{d^3 p}{(2\pi)^32E_{\bf p}}\int_0^tdx^0\int_0^{x_0}dy^0\int d^3 x\int d^3y \,  \rho(x) \, \rho(y) \, {\rm sin} \left[ p\cdot(x-y) \vphantom{\frac{}{}} \right]\in {\mathds R}.
    \end{split}
\end{equation}

\section{Decay in \textit{n} particles}
\label{sec:app_dec}
In this appendix we give a reminder in which sense the following formulas

\begin{equation}
    P_{\psi_i\to 0}(t_f,t_i)=\left|\vphantom{\frac{}{}} \langle 0|U(t_f,t_i)|\psi_i\rangle \right|^2, \qquad
     P_{\psi_i\to n}(t_f,t_i) = \int\frac{d^3 p_1 \cdots d^3 p_n}{(2\pi)^{3n} 2E_{{\bf p}_1}\cdots 2E_{{\bf p}_n}} \left| \vphantom{\frac{}{}}\langle {\bf p}_1\cdots{\bf p}_n|U(t_f,t_i)| \psi_i\rangle \right|^2,
\end{equation}
represent the probability that a system measured in the state $|\psi_i\rangle$ at the initial time $t_i$, is measured to be a $n$-particles state at time $t_f$, regardless of the details of the $n$-particle part of the wave function.
\\
That is the probability reads

\begin{equation}
    P_{\psi_i\to 0}(t_f,t_i)=\left|\vphantom{\frac{}{}} \langle 0|U(t_f,t_i)|\psi_i\rangle \right|^2, \qquad
     P_{\psi_i\to n}(t_f,t_i) = \int\frac{d^3 p_1 \cdots d^3 p_n}{(2\pi)^{3n} 2E_{{\bf p}_1}\cdots 2E_{{\bf p}_n}} \left| \vphantom{\frac{}{}}\langle {\bf p}_1\cdots{\bf p}_n|U(t_f,t_i)| \psi_i\rangle \right|^2,
\end{equation}
The amplitude, which gives the transition probability, in the Heisenberg picture reads

\begin{equation}
    a_{i\to f}=\langle \psi_f|U(t_f,t_i)|\psi_i\rangle,
\end{equation}
with an explicit mention of the final state $|\psi_f\rangle$, but no explicit concept of particle number.
\\
Assuming that the eigenstates of all observables at the time $t_f$ can be decomposed exclusively with the basis $|0\rangle$, $|{\bf p}_1\cdots{\bf p}_n\rangle$, the final state $ |\psi_f\rangle$ can be expressed as

\begin{equation}
    \begin{split}
    |\psi_f\rangle &= {\mathds 1}|\psi_f\rangle = \langle0|\psi_f\rangle|0\rangle +\sum_{n=1}^\infty \int\frac{d^3p_1\cdots d^3p_n}{(2\pi)^{3n}2E_{{\bf p}_1}\cdots2E_{{\bf p}_n}}\langle{\bf p}_1,\cdots, {\bf p}_n|\psi_f\rangle |{\bf p}_1,\cdots, {\bf p}_n\rangle \equiv \\
    &\equiv \psi_{f}^{(0)}|0\rangle + \sum_{n=1}^\infty \int\frac{d^3p_1\cdots d^3p_n}{(2\pi)^{3n}2E_{{\bf p}_1}\cdots2E_{{\bf p}_n}} \; \psi^{(n)}_{f}({\bf p}_1,\cdots, {\bf p}_n)|{\bf p}_1,\cdots, {\bf p}_n\rangle.
    \end{split}
\end{equation}
One has a clear understanding what it means to have an $\tilde{n}$-particle final state. The partial wave for all $n\neq\tilde{n}$ is zero, $\psi_{f}^{(n\neq\tilde{n})}=0$.

Making a similar decomposition for $U|\psi_i\rangle$, which is the time-dependent state in the Shr\"{o}dinger picture, one has

\begin{equation}
    U(t_f,t_i)|\psi_i\rangle = \sum_{n=0}^\infty \int\frac{d^3p_1\cdots d^3p_n}{(2\pi)^{3n}2E_{{\bf p}_1}\cdots2E_{{\bf p}_n}}\langle{\bf p}_1,\cdots, {\bf p}_n| U(t_f,t_i)|\psi_i\rangle |{\bf p}_1,\cdots, {\bf p}_n\rangle.
\end{equation}
Calling $|\psi_{i}^{(n)}(t_f,t_i)\rangle$ the $n$-particle partial wave, after normalizing it to $1$

\begin{equation}
    \begin{split}
        |\psi_{i}^{(n)}(t_f,t_i)\rangle &=\frac{1}{N^{(n)}_{\psi_i}(t_f,t_i)}\int\frac{d^3p_1\cdots d^3p_n}{(2\pi)^{3n}2E_{{\bf p}_1}\cdots2E_{{\bf p}_n}}\langle{\bf p}_1,\cdots, {\bf p}_n| U(t_f,t_i)|\psi_i\rangle |{\bf p}_1,\cdots, {\bf p}_n\rangle, \\
        \left( N^{(n)}_{\psi_i}(t_f,t_i) \right)^2 &= \int\frac{d^3p_1\cdots d^3p_n}{(2\pi)^{3n}2E_{{\bf p}_1}\cdots2E_{{\bf p}_n}}\left|\langle{\bf p}_1,\cdots, {\bf p}_n| U(t_f,t_i)|\psi_i\rangle \vphantom{\frac{}{}}\right|^2 = P_{\psi_i\to n}(t_f,t_i).
    \end{split}
\end{equation}
In other words

\begin{equation}
    U(t_f,t_i)|\psi_i\rangle = \sum_{n=0}^\infty \sqrt{P_{\psi_i\to n}(t_f,t_i)} \, |\psi_{i}^{(n)}(t_f,t_i)\rangle,
\end{equation}
that is $P_{\psi_i\to n}(t_f,t_i)$ is the largest probability to measure an $n$-particles state, which is realized if and only if $|\psi_f\rangle =|\psi_{i}^{(n)}(t_f,t_i)\rangle$.
\\
Another way to put it, $P_{\psi_i\to n}(t_f,t_i)$ is the probability to decay in $n$-particles in the sense that it follows the usual rules of probability decomposition, despite quantum mechanics revolving around amplitudes. To be more precise, being $|\psi^{(n)}_f\rangle$ a purely $n$-particles final state, the probability to measure it at time $t_f$ reads

\begin{equation}
   \left| \langle \psi^{(n)}_f|U(t_f,t_i)|\psi_i\rangle \vphantom{\frac{}{}}\right|^2 = P_{\psi_i\to n}(t_f,t_i) \left| \langle \psi^{(n)}_f|\psi_{i}^{(n)}(t_f,t_i)\rangle \vphantom{\frac{}{}}\right|^2.
\end{equation}
That is, the probability to decay into an $n$-particle state times the probability that the $n$-particles state is actually $|\psi^{(n)}_f\rangle$ (the overlap between the two states). This interpretation of $P_{\psi_i\to n}(t_f,t_i)$ is the only one compatible with both the regular factorization of constrained probabilities, and the superposition rules in the states of quantum mechanics.

\section{Coherent states}
\label{sec:app_coh}
Eigenvectors $|c\rangle$ of the destruction operators are called coherent states
\begin{equation}
    a|c\rangle = c |c\rangle.
\end{equation}
The coherent state itself can be built noting that, as a consequence of the commutation relations

\begin{equation}\label{comm_disc}
\begin{split}
    [a,a^\dagger] = 1, \quad \Rightarrow \quad  a|n\rangle  = n|n-1\rangle,
\end{split}
\end{equation}
the formal series 

\begin{equation}\label{unnorm_disc}
    |0\rangle + \sum_{n=1}^\infty \frac{c^n}{\sqrt{n!}} |n\rangle = \left[\sum_{n=0}^\infty \frac{\left( c\, a^\dagger \vphantom{\frac{}{}} \right)^n}{n!}\right]|0\rangle = e^{c\, a^\dagger}
|0\rangle, \qquad \Rightarrow \qquad a \left( e^{c\, a^\dagger}
|0\rangle \right) = c \left( e^{c\, a^\dagger}
|0\rangle \right),
\end{equation}
is an eigenvector of $a$ with with eigenvalue $c$, an arbitrary complex number. In order to be an eigenstate it must be normalized to $1$.

The normalization can be computed in a straightforward manner using~\eqref{unnorm_disc}

\begin{equation}
    \begin{split}
        \left| e^{c\, a^\dagger}
|0\rangle \right|^2 &= \left(\langle 0| e^{c^*\, a}
 \right)\left( e^{c\, a^\dagger}
|0\rangle \right) = \langle 0| \left[\sum_{n=0}^\infty \frac{\left( c^*\, a\vphantom{\frac{}{}} \right)^n}{n!}\right]\left( e^{c\, a^\dagger}
|0\rangle \right) = \langle 0|0\rangle \sum_{n=0}^\infty \frac{\left| \vphantom{\frac{}{}} \; c \; \right|^{2n}}{n!} = e^{\left| \vphantom{\frac{}{}} \; c \; \right|^{2}}.
    \end{split}
\end{equation}
Therefore the physical coherent state is given by
\begin{equation}
    |c\rangle = e^{-\frac{1}{2}|c|^2} \, e^{ c\, a^\dagger}|0\rangle.
\end{equation}
The generalization to the uncharged scalar field situation

\begin{equation}
    a_{\bf p}|\psi\rangle = -h({\bf p}) |\psi\rangle,
\end{equation}
is rather straightforward. One has to take into account that there are multiple creation-annihilation operators (for each momentum ${\bf p}$), that each annihilation operator has a different eigenvalue, and that the normalization is not to the identity but to a distribution

\begin{equation}
    [a_{\bf p}, a^\dagger_{\bf q}]= (2 \pi^3) \delta({\bf p}-{\bf q}).
\end{equation}
Instead of a straight multiplication of the creation operators times the eigenvalue, one needs the multiplication of the momentum integral of them

\begin{equation}
    \int\frac{d^3 p}{(2\pi)^3} \, \left( \vphantom{\frac{}{}} - h(\bf p)\right) \; a^\dagger_{\bf p}.  
\end{equation}
Therefore, without loss of generality, an eigenvector for each $a_{\bf p}$ corresponding to the eigenvalue $-h({\bf p})$ is

\begin{equation}\label{unnorm_con}
\begin{split}
    & |0\rangle + \left[\sum_{n=1}^\infty \frac{1}{n!} \int\frac{d^2p_1d^3p_2\cdots d^3 p_n}{(2\pi)^{3n}} \left( \vphantom{\frac{}{}} -h({\bf p}_1) a^\dagger_{{\bf p}_1} \right)\left( \vphantom{\frac{}{}} -h({\bf p}_2) a^\dagger_{{\bf p}_2} \right) \cdots \left( \vphantom{\frac{}{}} -h({\bf p}_n) a^\dagger_{{\bf p}_n} \right) \right]|0\rangle = \\
    &= \left[\sum_{n=0}^\infty \frac{1}{n!}\left( -\int\frac{d^3 p}{(2\pi)^3 }\, h({\bf p}) \, a^\dagger_{\bf p} \right)^n\right]|0\rangle = {\rm exp}\left\{-\int\frac{d^3 p}{(2\pi)^3} \, h({\bf p}) \, a_{\bf p}^\dagger \right\}
|0\rangle, \qquad \Rightarrow \\ \\
& \Rightarrow \qquad  a_{\bf p} \; {\rm exp}\left\{-\int\frac{d^3 p}{(2\pi)^3} \, h({\bf p}) \, a_{\bf p}^\dagger \right\}
|0\rangle  = -h({\bf p}) \; {\rm exp}\left\{ -\int\frac{d^3 p}{(2\pi)^3} \, h({\bf p}) \, a_{\bf p}^\dagger \right\}
|0\rangle .
\end{split}
\end{equation}
The normalization follows from the commutation relations in a very similar way as in the discrete case. Therefore

\begin{equation}
\begin{split}
    & |\psi\rangle \quad {\rm s.t.} \quad a_{\bf p}|\psi\rangle = -h({\bf p}) |\psi\rangle, \quad \Rightarrow\\
    & \qquad \Rightarrow \quad |\psi\rangle = {\rm exp}\left\{ -\frac{1}{2}\int\frac{d^3 p}{(2\pi)^3} \, \left|h({\bf p}) \right|^2 \right\}{\rm exp}\left\{ -\int\frac{d^3 p}{(2\pi)^3} \, h({\bf p}) \, a_{\bf p}^\dagger \right\}
|0\rangle.
\end{split}
\end{equation}

\section{Scattering states}
\label{sec:app_Smatrix}
Given the important role that the definition of the scattering states hold in this work, in this section there is a reminder of the most common concept.
Within the Heisenberg representation, the in and out states

\begin{equation}
    |S_{\rm in}\rangle, \qquad |S_{\rm out}\rangle,
\end{equation}
are states that in the distant past (for the in-states) and in the distant future (out-states) resemble some free states. In the Heisenberg picture, of course, the states do not depend on time, the former statements mean that the partial waves can be approximated by some free wave-function

\begin{equation}
    \langle{\bf p}_{1}\cdots {\bf p}_{n}|U(t,0)|S_{{\rm in} / {\rm out}}\rangle \simeq  e^{-i\sum_j E_{{\bf p}_j}t} \;\psi_{{\rm in} / {\rm out}}({\bf p}_1,\cdots{\bf p}_n),
\end{equation}
and that the approximation becomes more accurate the smaller(larger) time $t$ for the in(out) case. The basis $|{\bf p}_{1}\cdots {\bf p}_{n}\rangle$ is the regular one for the free state. This is built from interaction picture fields $\hat \phi_I(x)$ which coincide with the free fields at time $t=0$ and fulfill the free Klein-Gordon equation. The overlap between the states

\begin{equation}
    \langle S_{\rm out}|S_{\rm in}\rangle,
\end{equation}
has the immediate interpretation of the probability amplitude of a state (arbitrarily close to a free state in the distant past) that evolves into some final state (also free-like) in the distant future.

The simpler way to isolate such states is to modify the theory, having the interaction (and eventual counterterms) only for a limited time, from $t_-$ to $t_+>t_-$. In this way the states do not resemble free states in the distant past or future, they become free states. In the sense that, in the Shr\"odinger picture, they evolve according to the free propagator $U_0 $. Namely, for the in states

\begin{equation}\label{in_SP}
\begin{split}
    U(t<t_-,0)|S_{\rm in}\rangle &= U(t,t_-) \left[U(t_-,0)\vphantom{\frac{}{}}|S_{\rm in}\rangle \right] = U_0(t,t_-) \left[U(t_-,0)\vphantom{\frac{}{}}|S_{\rm in}\rangle \right] 
    \\
    &=\sum_{n=0}^\infty \int \prod_{j=1}^n \left\{\frac{d^3 p_j}{(2\pi)^3 2 E_{{\bf p}_j}}\right\} e^{-i\sum_{j}E_{{\bf p}_j}t} \; \psi_{\rm in}({\bf p}_1,\cdots{\bf p}_n)|{\bf p}_1,\cdots{\bf p}_n\rangle.
\end{split}
\end{equation}
For $t<t_-$ the system is free and the decomposition of the state at time $t_-$ in the partial wavefucntions is not ambiguous. The phase convention for the $\psi_{\rm in}$ is, clearly

\begin{equation}
    \langle {\bf p}_1,\cdots{\bf p}_n|U(t_-,0)|S_{\rm in}\rangle =  e^{-i\sum_{j}E_{{\bf p}_j}t_-}\psi_{\rm in}({\bf p}_1,\cdots{\bf p}_n),
\end{equation}
since the evolution is free $e^{-i {\hat H}_0 \, \Delta t}$ only before $t_-$ (and after $t_+$). The analogue holds for the out states

\begin{equation}\label{out_SP}
\begin{split}
    U(t>t_+,0)|S_{\rm out}\rangle &= U(t,t_+) \left[U(t_+,0)\vphantom{\frac{}{}}|S_{\rm out}\rangle \right] = U_0(t,t_+) \left[U(t_+,0)\vphantom{\frac{}{}}|S_{\rm out}\rangle \right] 
    \\
    &=\sum_{n=0}^\infty \int \prod_{j=1}^n \left\{\frac{d^3 p_j}{(2\pi)^3 2 E_{{\bf p}_j}}\right\}e^{-i\sum_{j}E_{{\bf p}_j}t} \; \psi_{\rm out}({\bf p}_1,\cdots{\bf p}_n)|{\bf p}_1,\cdots{\bf p}_n\rangle.
\end{split}
\end{equation}
The last formulas~\eqref{in_SP} and~\eqref{out_SP} can be inverted, in order write the states $|S_{{\rm in}/{\rm out}}\rangle$ according to something recognizable $|S_{{\rm in}/{\rm out}}\rangle = U^{-1}\left(\;\right.$ free-state decomposition $\left.\;\right)$. In this way the overlap reads

\begin{equation}\label{amp_SP}
    \begin{split}
        &\langle S_{\rm out}|S_{\rm in}\rangle = \sum_{m,n}\int \prod_{j=1}^n\prod_{k=1}^m \left\{\frac{d^3 \tilde{p}_j}{(2\pi)^3 2 E_{\tilde{\bf p}_j}}\frac{d^3 p_k}{(2\pi)^3 2 E_{{\bf p}_k}} e^{iE_{\tilde{\bf p}_j}t_f-iE_{{\bf p}_k}t_i}\right\} \\
        & \qquad\times \psi_{\rm out}^*(\tilde{\bf p}_1,\cdots \tilde{\bf p}_n)\psi_{\rm in}({\bf p}_1,\cdots {\bf p}_m) \; \langle\tilde{\bf p}_1,\cdots \tilde{\bf p}_n|U(t_f,0)U^{-1}(t_i,0)|{\bf p}_1,\cdots {\bf p}_m\rangle = \\
        &=\sum_{m,n}\int \prod_{j=1}^n\prod_{k=1}^m \left\{\frac{d^3 \tilde{p}_j}{(2\pi)^3 2 E_{\tilde{\bf p}_j}}\frac{d^3 p_k}{(2\pi)^3 2 E_{{\bf p}_k}} e^{iE_{\tilde{\bf p}_j}t_f-iE_{{\bf p}_k}t_i}\right\}\psi_{\rm out}^*(\tilde{\bf p}_1,\cdots \tilde{\bf p}_n)\psi_{\rm in}({\bf p}_1,\cdots {\bf p}_m) \\
        & \qquad \qquad \qquad \times\langle\tilde{\bf p}_1,\cdots \tilde{\bf p}_n|U(t_f,t_i)|{\bf p}_1,\cdots {\bf p}_m\rangle,
    \end{split}
\end{equation}
for any $t_f\ge t+$ and $t_i\le t_-$. The problem is reformulated in terms of the matrix elements of the propagator $U$.  This formula can be significantly simplified, removing the phases on the energies and removing the explicit dependence on  $t_f$ and $_i$. Instead of directly inverting~\eqref{in_SP} and~\eqref{out_SP}, first applying $U_0^{-1}$

\begin{equation}\label{in_out_IP}
\begin{split}
    & U_I(t<t_-,0)|S_{\rm out}\rangle = U_0^{-1}(t<t_-,0)U(t<t_-,0)|S_{\rm out}\rangle 
    \\
    &\quad = U_0^{-1}(t<t_-,0)\sum_{n=0}^\infty \int \prod_{j=1}^n \left\{\frac{d^3 p_j}{(2\pi)^3 2 E_{{\bf p}_j}}\right\}e^{-i\sum_{j}E_{{\bf p}_j}t} \; \psi_{\rm in}({\bf p}_1,\cdots{\bf p}_n)|{\bf p}_1,\cdots{\bf p}_n\rangle \\
    &\qquad =\sum_{n=0}^\infty \int \prod_{j=1}^n \left\{\frac{d^3 p_j}{(2\pi)^3 2 E_{{\bf p}_j}}\right\} \; \psi_{\rm in}({\bf p}_1,\cdots{\bf p}_n)|{\bf p}_1,\cdots{\bf p}_n\rangle , \\ \\
    & U_I(t>t_+,0)|S_{\rm out}\rangle = U_0^{-1}(t>t_+,0)U(t>t_+,0)|S_{\rm out}\rangle 
    \\
    &\quad = U_0^{-1}(t>t_+,0)\sum_{n=0}^\infty \int \prod_{j=1}^n \left\{\frac{d^3 p_j}{(2\pi)^3 2 E_{{\bf p}_j}}\right\}e^{-i\sum_{j}E_{{\bf p}_j}t} \; \psi_{\rm out}({\bf p}_1,\cdots{\bf p}_n)|{\bf p}_1,\cdots{\bf p}_n\rangle \\
    &\qquad =\sum_{n=0}^\infty \int \prod_{j=1}^n \left\{\frac{d^3 p_j}{(2\pi)^3 2 E_{{\bf p}_j}}\right\} \; \psi_{\rm out}({\bf p}_1,\cdots{\bf p}_n)|{\bf p}_1,\cdots{\bf p}_n\rangle.
\end{split}
\end{equation}
Then, inverting these formulas with the interaction picture propagator on the left hand side, and following the same passages that brought~\eqref{amp_SP}, one has the alternative formula

\begin{equation}\label{amp_IP}
    \begin{split}
        &\langle S_{\rm out}|S_{\rm in}\rangle = \sum_{m,n}\int \prod_{j=1}^n\prod_{k=1}^m \left\{\frac{d^3 \tilde{p}_j}{(2\pi)^3 2 E_{\tilde{\bf p}_j}}\frac{d^3 p_k}{(2\pi)^3 2 E_{{\bf p}_k}} \right\}\psi_{\rm out}^*(\tilde{\bf p}_1,\cdots \tilde{\bf p}_n)\psi_{\rm in}({\bf p}_1,\cdots {\bf p}_m) \\
        & \qquad \qquad \qquad \qquad \qquad \times\langle\tilde{\bf p}_1,\cdots \tilde{\bf p}_n|U_I(t_+,t_-)|{\bf p}_1,\cdots {\bf p}_m\rangle.
    \end{split}
\end{equation}
The problem is reduced to the calculation of the matrix elements of the interaction picture propagator $U_I= U_0^\dagger U$. The explicit dependence on $t_-$ and $t_+$ is still there. The hope is that, having the regular (unaltered) interaction for the bulk of $\{t_-,t_+\}$ plus some reconnection to zero close to the border of this integral, if both the limits

\begin{equation}
    \lim_{t_-\to-\infty}\lim_{t_+\to \infty}\langle\tilde{\bf p}_1,\cdots \tilde{\bf p}_n|U_I(t_+,t_-)|{\bf p}_1,\cdots {\bf p}_m\rangle,  \qquad \lim_{t_+\to\infty}\lim_{t_-\to -\infty}\langle\tilde{\bf p}_1,\cdots \tilde{\bf p}_n|U_I(t_+,t_-)|{\bf p}_1,\cdots {\bf p}_m\rangle,
\end{equation}
exist and are equal, then the results of the full theory will be recovered. These limits are important since the $d^4x$ integrals produce the four-momentum conservation deltas at the vertices in the Feynman rules. It is immediate to notice that, within this framework, stable bound states like the Hydrogen atom in its fundamental state cannot be treated. One needs to use some effective Lagrangian that includes all stable states in the $\hat H_0$.

Besides this, in order to check if this procedure of switching the interaction reproduces the results of the full theory, at least regarding the stable states, an alternative treatment is necessary. In axiomatic field theory~\cite{Ticciati:1999qp} one assumes (or proves, if possible) that the four-momentum operators $\hat P^\mu$ have a (physical) vacuum eigenstate\footnote{The Hamiltonian operator in quantum field theory usually has to be just bounded from below. A constant (and finite) counterterm is necessary, in general, to ensure that the minimum is exactly zero. }
 $|0\rangle_P$
\begin{equation}\label{phys_vac}
    \hat P^\mu|0\rangle_P=0, \qquad {}_P\langle 0|0\rangle_P =1,
\end{equation}
which is orthogonal to all the other physical states. In addition it is required that a subspace of one-particle states is spanned by the formal one-particle states

\begin{equation}
    \hat P^\mu|{\bf p} \rangle=p^\mu|{\bf p} \rangle, \quad \langle{\bf p}|{\bf q}\rangle =(2 \pi)^3 2E_{\bf p}\delta^3({\bf p}- {\bf q}),\quad {}_P\langle 0|{\bf p}\rangle = 0,
\end{equation}
for some physical mass $m\to p^0=E_{\bf p}=\sqrt{m^2 +{\bf p}\cdot{\bf p}}$. In the case of scalar particles, there must be a scalar field\footnote{Usually it is necessary to renormalize the fields, at least remove the vacuum expectation value and rescale the field with a positive constant. See, eg, the book~\cite{Ticciati:1999qp}.} $\hat\phi^\prime$, which does not have to be fundamental, such that

\begin{equation}\label{1_part_phi}
    {}_P\langle 0|\hat \phi^\prime(x)|0\rangle_P=0, \qquad   {}_P\langle0|\hat \phi^\prime(x)|{\bf p}\rangle = e^{-ip\cdot x}.
\end{equation}
These are the generalization to the interacting theory of the analogous formulas fulfilled by free fields. The one-particle physical states can be built just like in the free case

\begin{equation}
    |\psi_{(1)}\rangle = \int \frac{d^3 p}{(2\pi)^3 2E_{\bf p}} \; \psi_{(1)}({\bf p}) \, |{\bf p} \rangle, \qquad \langle \psi_{(1)}|\psi_{(1)} \rangle =1,
\end{equation}
for some wavefunction (in the Heisenberg representation) $\psi_{(1)}({\bf p})$. These states correspond to the free wave-functions (in the Shr\"odinger representation)

\begin{equation}
    \psi_{(1)}(t,{\bf x}) = \int \frac{d^3 p}{(2\pi)^3 2E_{\bf p}} \; e^{-iE_{\bf p}t + i {\bf p}\cdot{\bf x}}\,\psi_{(1)}({\bf p}).
\end{equation}
Following~\cite{Ticciati:1999qp}, it is possible then to build an operator that behaves like a creation operator of a physical one-particle state $\hat \phi^\prime[\psi_{(1)}]$

\begin{equation}
   \hat \phi^\prime[\psi_{(1)}](t) = i\int d^3 x \left[ \hat \phi^\prime(x) \,\partial_t \psi_{(1)}(x)- \psi_{(1)}(x) \, \partial_t\hat \phi^\prime(x) \right].
\end{equation}
Thanks to~\eqref{1_part_phi} it is immediate to prove that

\begin{equation}
    \langle{\bf p}| \; \hat \phi^\prime[\psi_{(1)}] (t) \, |0\rangle_P = \psi_{(1)}(p),\qquad {}_P\langle 0|\; \hat \phi^\prime[\psi_{(1)}](t)\, |{\bf p}\rangle = 0.
\end{equation}
Differently from the free case, the overlap of the ``one-particle'' state $\hat \phi^\prime[\psi_{(1)}]|0\rangle_P$ with a generic state $|\Psi\rangle$ is not the overlap of the one-particle partial wave  $\langle{\bf p}|\Psi\rangle$ with $\psi_{(1)}(p)$. However it can be shown that this expected overlap is the only remaining term both in the $t\to \infty$ limit, and in the $t\to -\infty$ one. The asymptotic multiparticle-states are obtained as the limits

 \begin{equation}
     |\psi_{(1)}^a \, \psi_{(1)}^b, \cdots\rangle_{{\rm in}/{\rm out}} = \lim_{t\to \pm\infty}\left\{\left(\hat \phi^\prime[\psi_{(1)}^a](t) \; \hat \phi^\prime[\psi_{(1)}^b](t)\cdots \right)|0\rangle_P\right\}.
 \end{equation} 
Either one accept this as a definition, possibly recurring to some intuitive localization arguments as in~\cite{Coleman:2018mew,Ticciati:1999qp}, otherwise it is necessary to delve deeply in the mathematical structure of axiomatic field theory~\cite{Bogolubov:1990ask}. The results of scattering theory obtained from the switch on approach can be recovered, for instance the Feynmann rules. There is the additional advantage to include in the scattering theory the stable bound states like the hydrogen atom.

In any case this approach cannot be used in this particular case. From the very start, the vacuum state in~\eqref{phys_vac} does not exist. The Hamiltonian~\eqref{hamiltonian_0} is bounded from below. It is not difficult to set the minimum to $0$ as required. It is enough a constant off-set in the Lagrangian (not in the Lagrangian density). However the minimum of the Hamiltonian is an eigenstate only of the Hamiltonian operator itself. It is not an eigenstate of the momentum $\hat{\bf P}(t)$ at any time (not even $t=0$), as shown in a appendix~\ref{sec:app_time_dep}: The energy vacuum $|\Omega(t)\rangle$ is a coherent state of the momentum eigenstates at time $t=0$ as in~\eqref{Omega_t}. In particular it is time-independent for a time independent source $\rho({\bf x})$. Since the operators $b(t,{\bf p})$ can be written as $a_{\bf p}$ plus a function, just like the $a(t,{\bf p})$, it is always possible to decompose $b(t,{\bf p}) = a(t,{\bf p})$ plus a function, repeat the passages and prove that minimum of the energy is always a coherent state of the momentum eigenstates, regardless of the time $t$ and it is never a momentum eigenstate, since the difference $b(t,{\bf p})-a(t,{\bf p})$ is never vanishing for a time independent source $\rho({\bf x})$.

It is necessary then to abandon the idea of asymptotic free states. The $t-$dependent basis $|{\bf p}_1,{\bf p}_2, \cdots {\bf p}_n\rangle_{t} $ in~\eqref{t_states} can be used to define the probability density to find $n$ particles with momentum ${\bf p}_1,{\bf p}_2\cdots {\bf p}_n$ with the usual algebra of free wave-functions\footnote{It is necessary to add energy phases as in~\eqref{scattering_time_dep}, as it can be immediately seen by the comparison with the formulas in the free case, if one wants to compute the probability density of the positions, which correspond to the $i\partial_{({\bf p})}$ derivatives.}. This is not very far from the treatment of unstable particles. They cannot be defined as asymptotic states in general. In the limit of infinite time the partial wave of a Lambda particle goes to zero. However it makes sense to talk about a measured unstable particle from the practice in particle physics. Their wave-function is usually written in the form, eg for a Lambda

\begin{equation}
    |\psi_\Lambda\rangle = \int \frac{d^3 p}{(2 \pi)^3}\psi_\Lambda({\bf p}) |{\bf p}\rangle_\Lambda,
\end{equation}
the $|{\bf p}\rangle_\Lambda$ are eigenstates of the momentum operator, and of the other observables fixed by the Lambda state (baryon number electric charge etc.). Usually the basis $|{\bf p}\rangle_\Lambda$ itself is assumed to be decomposable in terms of four-momentum eigenstates, according to the spectral decomposition $a_\Lambda(m)$

\begin{equation}
    |{\bf p}\rangle_\Lambda = \int_{m_{th}}^\infty dm \; a_\Lambda(m) |m, {\bf {\bf p}}\rangle_\Lambda,
\end{equation}
in which the mass parameter $m$ goes from minimum threshold $m_{th}$ and the $|m, {\bf {\bf p}}\rangle_\Lambda$ basis has the obvious energy $\sqrt{m^2 +{\bf p}\cdot{\bf p}}$. This passage is clearly inadequate in the theory analyzed in this work, since there is no basis of simultaneous eigenstates of both the energy and the free-momentum operators. However the rest of the analogy holds. It is important to note that in both cases, unstable states in the standard model and the more general $t-$dependent basis decomposition in this work, the wave-functions are not behaving as free ones, not even for an arbitrarily small time interval. For instance the velocity (the expectation value of the ratio of the momentum over the energy) is not constant.

That said, it is possible to compute, as has been done in this work, the matrix elements of the interaction picture propagator $U_{I}(t_f,t_i)$ and make the limits $t_f\to \infty$ and $t_i\to-\infty$. Despite the lack of asynmptotic states, in this way it is possible to check the asymptotic probabilities to measure a transition from one recognized state to another one.

\bibliographystyle{apsrev4-2.bst}
\bibliography{vacuum}

\end{document}